\documentclass[remotesensing,article,accept,pdftex,moreauthors]{Definitions/mdpi} 
\usepackage[htt]{hyphenat}
\usepackage[normalem]{ulem}

\newcommand{\gaia}{\textit{Gaia}}
\newcommand{\Gaia}{\textit{Gaia}}
\newcommand{\gdr}{\gaia~DR}
\newcommand{\gedr}{\gaia~EDR3}
\newcommand\mnras{Mon. Not. R. Astron. Soc.}             
\newcommand\aap{Astron. Astrophys.}                
\newcommand\aj{Astron. J.}                   
\newcommand\araa{Annu. Rev. Astron. Astrophys.}             
\newcommand\pasp{Publ. Astron. Soc. Pac.}               
\newcommand\apj{Astrophys. J.}                 

\firstpage{1} 
\pubvolume{1}
\issuenum{1}
\articlenumber{0}
\pubyear{2023}
\copyrightyear{2023}
\externaleditor{Academic Editor: Firstname Lastname}
\datereceived{31 January 2023} 
\daterevised{22 March 2023} 
\dateaccepted{ } 
\datepublished{ } 
\hreflink{https://doi.org/} 

\Title{Photometric Catalogue for Space and Ground Night-Time Remote-Sensing Calibration: RGB Synthetic Photometry from {\gdr3} Spectrophotometry}
\TitleCitation{Photometric Catalogue for Space and Ground Night-Time Remote-Sensing Calibration: RGB Synthetic Photometry from {\gdr3} Spectrophotometry}

\Author{Josep Manel Carrasco$^{1,2,3,}$*, Nicolas Cardiel $^{4,5}$, Eduard Masana $^{1,2,3}$, Jaime Zamorano $^{4,5}$, Sergio Pascual $^{4,5}$, Alejandro S\'{a}nchez de Miguel $^{4,5,6}$, Rafael González $^{4}$ and Jaime Izquierdo $^{4}$
}

\AuthorCitation{{Carrasco, J.M.;} Cardiel, N.; Masana, E.; Zamorano, J.; Pascual, S.; S\'{a}nchez de Miguel, A.; González, R.; Izquierdo, J.}

\address{%
$^{1}$ \quad Institut de Ci\`{e}ncies del Cosmos (ICCUB), Universitat  de  Barcelona  (IEEC-UB), Mart\'{i} i  Franqu\`{e}s  1, \mbox{08028~Barcelona, Spain}; emasana@fqa.ub.edu\\
$^{2}$ \quad Departament de Física Qu\`antica i Astrof\'isica (FQA), Universitat de Barcelona (UB), Mart\'i i
Franqu\`es 1, 08028~Barcelona Barcelona, Spain\\
$^{3}$ \quad Institut d'Estudis Espacials de Catalunya (IEEC), c. Gran Capit\`a, 2-4, 08034~Barcelona, Spain\\
$^{4}$ \quad Departamento de Física de la Tierra y Astrof\'{\i}sica, Facultad de CC.\ F\'{\i}sicas, Universidad Complutense de Madrid, Plaza de las Ciencias~1, 28040~Madrid, Spain; cardiel@ucm.es (N.C.); jzamorano@fis.ucm.es (J.M.); sergiopr@fis.ucm.es (S.P.); alejasan@ucm.es (A.S.d.M.); rafael08@ucm.es (R.G.); jiz@astrax.fis.ucm.es (J.I.)\\
$^{5}$ \quad Instituto de F\'{\i}sica de Part\'{\i}culas y del Cosmos, IPARCOS, Facultad de CC.\ F\'{\i}sicas, Universidad Complutense de Madrid, Plaza de las Ciencias~1, 28040~Madrid, Spain
\\
$^{6}$ \quad Environment and Sustainability Institute, University of Exeter, Penryn, Cornwall TR10 9FE, UK
}

\corres{Correspondence: carrasco@fqa.ub.edu }

 \abstract
   {Recent works have made strong efforts  to produce standardised photometry in RGB bands. For this purpose, we carefully defined the transmissivity curves of RGB bands and defined a set of standard sources   using the photometric information present in {\gedr}.
   This work aims  not only to significantly increase the number and accuracy of RGB standards but also to provide, for the first time, reliable uncertainty estimates  using the BP and RP spectrophotometry published in {\gdr3} instead of their integrated photometry to predict RGB photometry. Furthermore, this method allows including calibrated sources regardless of how they are affected by extinction, which was a major shortcoming of previous work. 
The RGB photometry is synthesised from the {\gaia} BP and RP low-resolution spectra by directly using their set of coefficients multiplied with some basis functions provided in the {\gaia} catalogue for all sources published in {\gdr3}. The output synthetic magnitudes are compared with the previous catalogue of RGB standards available.
}


\begin{document}

\section{Introduction}
\label{sec1}

RGB photometry has been increasingly used in recent decades for amateur and professional astronomical studies due to the high-quality and economically accessible digital cameras. In~recent works~\cite{Cardiel2021Def,Cardiel2021}, a~strong effort was made to produce a standardised system for RGB photometry to enhance the quality of the studies to be performed with these kind of devices. This is relevant not only  for ground measurements but also for satellite observations as~stars are also being used to calibrate night-time remote-sensing platforms, such as the images taken from the International Space Station (ISS)~\cite{de2021colour} and the Suomi North Polar Partnership and NOAA-20, Visible Infrared Imaging Radiometer Suite Day Night Band (VIIRS)~\cite{fulbright2015suomi,wilson2020intercomparison}.

Ref.~\cite{Cardiel2021Def} defined the transmissivity passbands for RGB filters derived from 28 different types of cameras analysed by~\cite{Jiang2013}. Ref.~\cite{Cardiel2021Def} also established  standardised synthetic RGB photometry for a set of 1346 bright stars belonging to the Bright Star Catalogue~\cite{Hoffleit1964}. 
  This small set of standards was expanded in number (about 15 million sources) by~{\cite{Cardiel2021}}(hereafter, C21) 
  using photometric transformations derived from integrated photometry in {\gedr} \cite{Brown2021,Riello2021}. 
This work aims to further expand the quality and number of sources in the sky with known RGB photometry that could be used as standards. For~that purpose, we use synthetic photometry derived from the {\gdr3} low-resolution spectra \citep{Carrasco2021,DeAngeli2022,Montegriffo2022ExtCal}. 

\begin{figure}[H]
    \includegraphics[width=1.0\columnwidth]{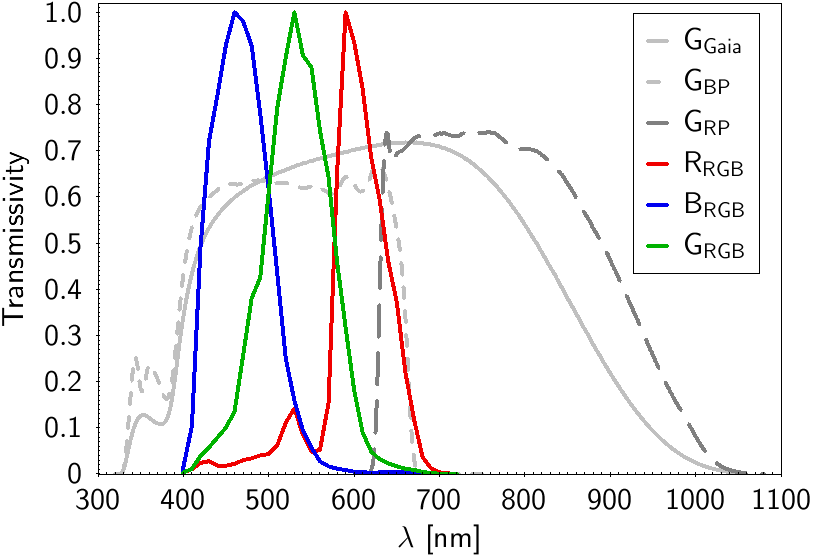}
    \caption{Comparison of RGB transmissivity curves (in colour lines) used in this work, extracted from~{\cite{Cardiel2021Def}} with~the {\gdr3} transmissivity curves by~{\cite{Riello2021}} (in grey). In~order to distinguish between the $G$ passband from the RGB system and the $G$ passband from the {\gaia} system, we add 'RGB' as a subscript to the first and 'Gaia' as a subscript to the latter.}
    \label{fig:transmissivity}
\end{figure}

In {\gdr3} \citep{Vallenari2022}, a set of 220 million sources were released together with their blue (BP) and red (RP) low-resolution spectra.
When compared with RGB passbands (Figure~\ref{fig:transmissivity}), we can see that RGB passbands mostly cover only the wavelength range covered by BP instrument (In order to avoid confusion, specifically with the $G$ band, in this paper, we used  $G_{\rm Gaia}$, $G_{\rm BP}$ and $G_{\rm RP}$ to refer to {\gaia} magnitudes and $R_{\rm RGB}$, $G_{\rm RGB}$ and $B_{\rm RGB}$ for magnitudes in the RGB system.). 
 
 It is known \citep{DeAngeli2022,Montegriffo2022ExtCal,Montegriffo2022SynthPhot} that the BP instrument has more calibration issues present than RP. Fortunately, most of these issues are assigned to difficulties in the ultraviolet region ($\lambda<400$~nm), where the {\gaia} response decreases abruptly, and the amount of standards with enough flux in that range diminishes substantially. As~none of the RGB passbands extends to so short wavelengths, we can still use {\gaia} spectrophotometry to derive synthetic RGB photometry from~them.

We describe, in Section~\ref{sec:methodology}, the methodology used to derive the synthetic photometry from {\gaia} spectrophometry, producing the synthetic photometry for all sources with BP and RP spectra (hereafter, XP is used to refer to BP or RP spectra indistinctly) present in {\gdr3} and not flagged as variable in the catalogue.
In Section~\ref{sec:comparison_with_C21}, we compare the samples of sources in this work with the ones used by C21, comparing also the RGB magnitudes obtained in both studies. In~Section~\ref{sec:validity_C21_polynomial_calibration}, we study the validity of the polynomials defined in C21 for those sources in {\gdr3} without XP spectra available. Section~\ref{sec:rgbloom} explains how to access the final catalogue of RGB magnitudes created in this work (in form of an online table and through a Python code  called {\sc rgbloom}. Finally, we present a brief summary of the results and conclusions of this work in Section~\ref{sec:conclusions}. 

\section{RGB Catalogue~Construction}
\label{sec:methodology}

{\gaia} \citep{Prusti2016} is a survey mission, which  allows homogeneously observing all sources in the sky up to magnitude 21. In addition to the~3D positions, motions and broad band photometry, {\gaia} also provides spectrophotometry for 219 million sources \citep{Carrasco2021,DeAngeli2022,Montegriffo2022ExtCal}. The~astrophysical information contained in these spectra  allows the study of the physical properties of the observed sources to analyse, for instance, their spectral features as performed in~\cite{Weiler2023}. 

The homogeneity of the spectrophotometry over all sky positions also makes {\gaia} a  suitable mission to be used as a reference to establish a catalogue of standard sources in any photometric system covering the optical range. This possibility to derive synthetic photometry from the {\gaia} spectrophotometry has been previously explored by several works \citep{Carrasco2017,Weiler2020,DeAngeli2022,Montegriffo2022SynthPhot}. We use here the same methodology (briefly described in the following paragraphs) to derive the synthetic photometry in RGB bands from XP {\gdr3} spectrophotometry. Notice that, as~recommended by~\cite{DeAngeli2022}, we do not use, in this work, the sampled spectra to derive the synthetic photometry; instead,~we  directly work with the source coefficients to obtain our estimations of the RGB~magnitudes.

From the externally calibrated {\gaia} mean spectra, $f_s(\lambda)$, for~a given source, $s$, we can derive its synthetic integrated flux ($F_{sj}$) in a given passband, $j$, with~transmissivity equal to $T_j(\lambda)$ by~deriving the following integral expression with wavelength $\lambda$:
\begin{equation}
\label{eq:integrals}
F_{sj}=\frac{\int f_s(\lambda) \cdot T_j(\lambda) \cdot  \lambda\ d\lambda}{\int T_j(\lambda) \cdot  \left(\frac{c}{\lambda}\right)\ d\lambda} \; ,
\end{equation}

\noindent where the ABMAG system is considered for the zeropoint \citep{fukugita1996,Bessell2005,Sirianni2005} with $c$ as the speed of~light.

The {\gaia} externally calibrated mean spectra $f_s(\lambda)$ is described as the weighted sum of the BP and RP contributions, $f^{\rm BP}_{s}(\lambda)$ and $f^{\rm RP}_{s}(\lambda)$, respectively:
\begin{equation}
\label{eq:weights}
f_s(\lambda)=w^{\rm BP}(\lambda)\cdot f^{\rm BP}_{s}(\lambda) + w^{\rm RP}(\lambda)\cdot f^{\rm RP}_{s}(\lambda) \; ,
\end{equation}

\noindent with $w^{\rm BP}(\lambda)$ and $w^{\rm RP}(\lambda)$ as their weighted contributions to the total flux of each wavelength from the XP instruments.

In its turn, $f^{\rm BP}_{s}(\lambda)$ and  $f^{\rm RP}_{s}(\lambda)$ are described as a set of $N$ coefficients, $b_{s n}$, multiplied by a set of basis functions, $\varphi_n$:
\begin{equation}
\label{eq:sourcebasisfunctions}
f^{\rm XP}_{s}(\lambda) = \sum\limits_{n=1}^{N}\, b^{\rm XP}_{s n} \cdot \varphi^{\rm XP}_n(\lambda) \; ,
\end{equation}

The set of basis functions, $\varphi^{\rm XP}_n(\lambda)$, and~weights, $w^{\rm XP}(\lambda)$, were published together with {\gdr3} and are available on this webpage:
 \href{https://www.cosmos.esa.int/web/gaia/dr3-xpmergexpsampling}{https://www.cosmos.esa.int/web/gaia/dr3-xpmergexpsampling}, accessed on 12 July 2022.

Combining Equations~(\ref{eq:integrals})--(\ref{eq:sourcebasisfunctions}), we can derive the synthetic flux, $F_{sj}$, in~a given passband $j$ as:
\begin{eqnarray}
F_{sj}&=&\sum\limits_{n=1}^{N}\, b^{\rm BP}_{s n} \cdot
\left(\frac{\int w^{BP}(\lambda) \cdot  \varphi^{\rm BP}_n(\lambda) \cdot T_j(\lambda) \cdot  \lambda\ d\lambda}{\int T_j(\lambda) \cdot  \left(\frac{c}{\lambda}\right)\ d\lambda} \right) \; +\nonumber \\
 &+&\sum\limits_{n=1}^{N}\, b^{\rm RP}_{s n} \cdot
\left(\frac{\int w^{RP}(\lambda) \cdot  \varphi^{\rm RP}_n(\lambda) \cdot T_j(\lambda) \cdot  \lambda\ d\lambda}{\int T_j(\lambda) \cdot  \left(\frac{c}{\lambda}\right)\ d\lambda} \right) \; \equiv \nonumber \\
&\equiv&\sum\limits_{n=1}^{N}\, b^{\rm BP}_{s n} \cdot X^{\rm BP}_{nj} + \sum\limits_{n=1}^{N}\, b^{\rm RP}_{s n} \cdot X^{\rm RP}_{nj} \; ,
\end{eqnarray}

\noindent where the $X^{\rm XP}_{nj}$ terms do not depend on the source and can be derived and stored previously for any passbands and then applied to any required source using their $b_{sn}^{\rm XP}$ coefficients without~the need to recompute $X^{\rm XP}_n$ terms~again.

These integrated fluxes can be also expressed in terms of ABMAG system magnitudes, $m_{sj}$, by~using the flux in that passband measured for an input flux of $3.631\times 10^{-23}$~W/Hz/m$^2$. 
We name that flux as ${\rm ZP}_j$. Thus, the magnitude can be expressed as:
\begin{equation}
\label{eq:magnitude}
 m_{sj}=-2.5\log \left( \frac{F_{sj}}{{\rm ZP}_j} \right) \; .
\end{equation}

From the covariance matrix, $C^{\rm XP}_{nm}$, assigned to the source coefficients, we can also derive the uncertainty in the derived magnitudes as:
\begin{equation}
    \sigma_{sj}=\sqrt{A^{\rm BP}_j+A^{\rm RP}_j}\; ,
\end{equation}

\noindent where $A^{\rm XP}_j$ can be derived using:
\begin{equation}
    A^{\rm XP}_j=\sum\limits_{n=1}^{N}\, X^{\rm XP}_{nj} \cdot \left(\sum\limits_{m=1}^{N}\, C^{\rm XP}_{nm} \cdot X^{\rm XP}_{mj} \right) \; .
\end{equation}

\begin{figure}[H]
    \includegraphics[width=0.49\textwidth]{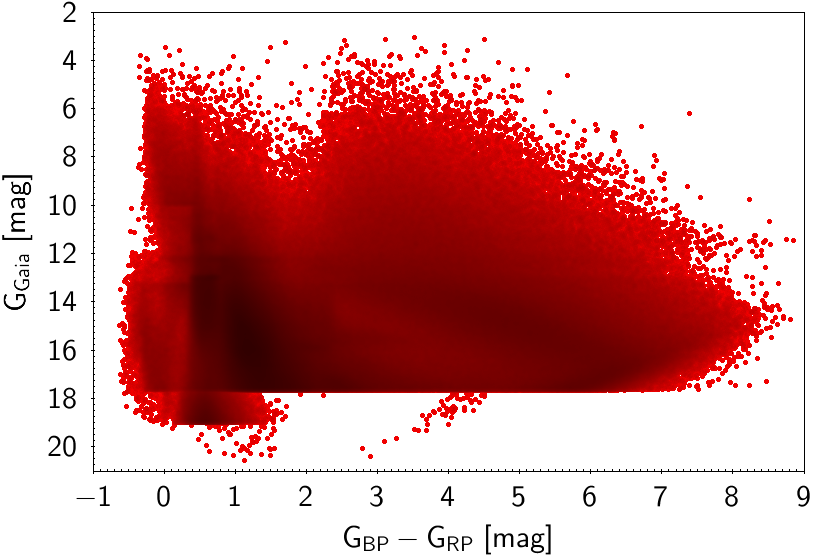}
    \includegraphics[width=0.49\textwidth]{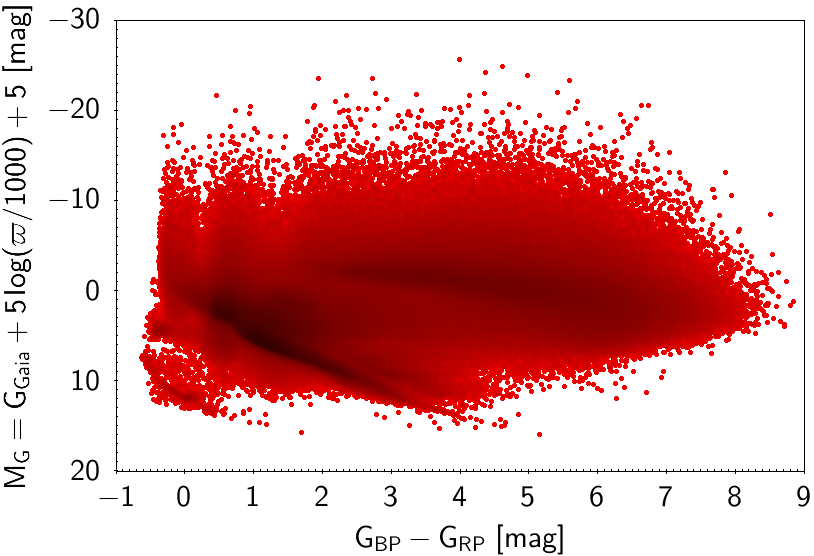}

    \caption{Magnitude-colour diagrams for sources with \texttt{phot\_variable\_flag='VARIABLE'} (excluded from our study) with continuous XP spectra in \emph{Gaia} DR3 using the apparent (left) and absolute (right) magnitude.}
    \label{fig:CMD_variables}
\end{figure}

\begin{figure}[H]
    \includegraphics[width=0.49\textwidth]{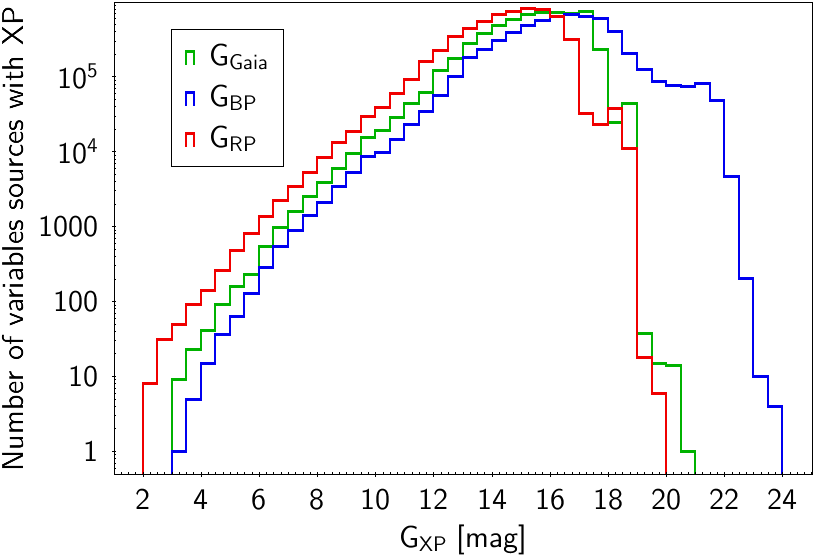}
    \includegraphics[width=0.49\textwidth]{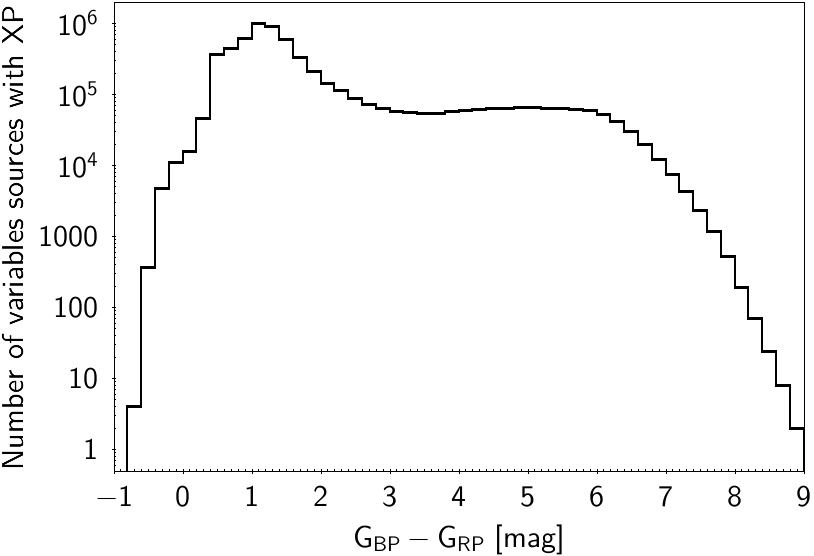}
    \caption{$G_{\rm Gaia}$ magnitude (left) and $G_{\rm BP}-G_{\rm RP}$ (right) histograms for sources with \texttt{phot\_variable\_flag='VARIABLE'} (excluded from our study) with continuous XP spectra in \emph{Gaia} DR3.}
    \label{fig:Histogram_variables}
\end{figure}

{\gdr3} published a total of 219,197,643 sources (see De Angeli et al. \cite{DeAngeli2022}) with XP continuous spectra. From~these, a~total of 6,093,025 sources have the flag \texttt{phot\_variable\_flag='VARIABLE'}.
We show the variable sources subtracted from the sample in colour-magnitude diagrams (Figure~\ref{fig:CMD_variables}) as well as their $G_{\rm Gaia}$ and $G_{\rm BP}-G_{\rm RP}$ histograms (Figure~\ref{fig:Histogram_variables}).

\unskip

These variable sources cannot be considered here for our purposes of establishing a set of reliable RGB
standards. We also aimed to select only sources with \mbox{\texttt{phot\_variable\_flag=}} \texttt{'CONSTANT'}
; however, no sources with {\gdr3} XP spectra have this flag value, as only sources with
\texttt{phot\_variable\_flag='NOT\_AVAILABLE'} and \texttt{'VARIABLE'} are present in the catalogue. This is because {\gdr3} is still an intermediate release and because the mission has not yet finished to be completely sure that a given source is really constant. Future {\gaia} releases could identify some sources finally included in the catalogue as variables when considering a larger time interval covering the full mission. Although other types of sources (quasars, galaxies, crowded fields, etc.) were not  suitable to be used as RGB standards, we decided to keep them in the catalogue in order to allow a wider range of applications for this catalogue, and~only explicit variable sources identified by the {\gaia} catalogue were excluded from it. With~the large density of sources available in the new catalogue, a~posterior outlier identification, based on statistical analysis, can be conducted in order to exclude them from the calibration procedure.

Thus, we derived the RGB synthetic magnitudes for all 213,104,618 sources with \texttt{phot\_variable\_flag$\neq$'VARIABLE'}. After~removing 40,616~sources with unreliable RGB magnitude estimates, a~final sample of 213,064,002~objects was created---hereafter, the 200 M sample. In~a small subsample of this collection, the RGB magnitude estimates are missing in some of the 3~bands (159,456, 14,546 and 486~objects in $B_{\rm RGB}$, $G_{\rm RGB}$ and $R_{\rm RGB}$, respectively). This is likely due to very low signal in these wavelength ranges present in the {\Gaia} spectrophotometry. We decided to keep the affected sources because, in all these cases, a magnitude estimate is available in at least one of the RGB~bands.

As the number of variable sources represents less than 3\% of the total number of {\gdr3} sources with XP data, the~general plots included in~\cite{DeAngeli2022} to describe {\gdr3} XP data, can also be used here to describe the main characteristics of our catalogue of RGB standards. The~only missing sources in our sample are the ones represented in \mbox{Figures~\ref{fig:CMD_variables} and~\ref{fig:Histogram_variables}}.

\section{Comparison with the C21~Sample}
\label{sec:comparison_with_C21}

The new 200 M sample exhibits clear advantages over its predecessor published by C21. The~most obvious advantage is the fact that the RGB magnitude estimates are directly computed from the source spectrum without the need to employ any approximate calibration, nor introducing constraints on the source colour or extinction. In~this section, we provide a more detailed description of the benefits of using the new 200 M sample provided with this~paper. 

\subsection{Number of Calibrated~Sources}

\begin{figure}[H]
    \includegraphics[width=0.8\textwidth]{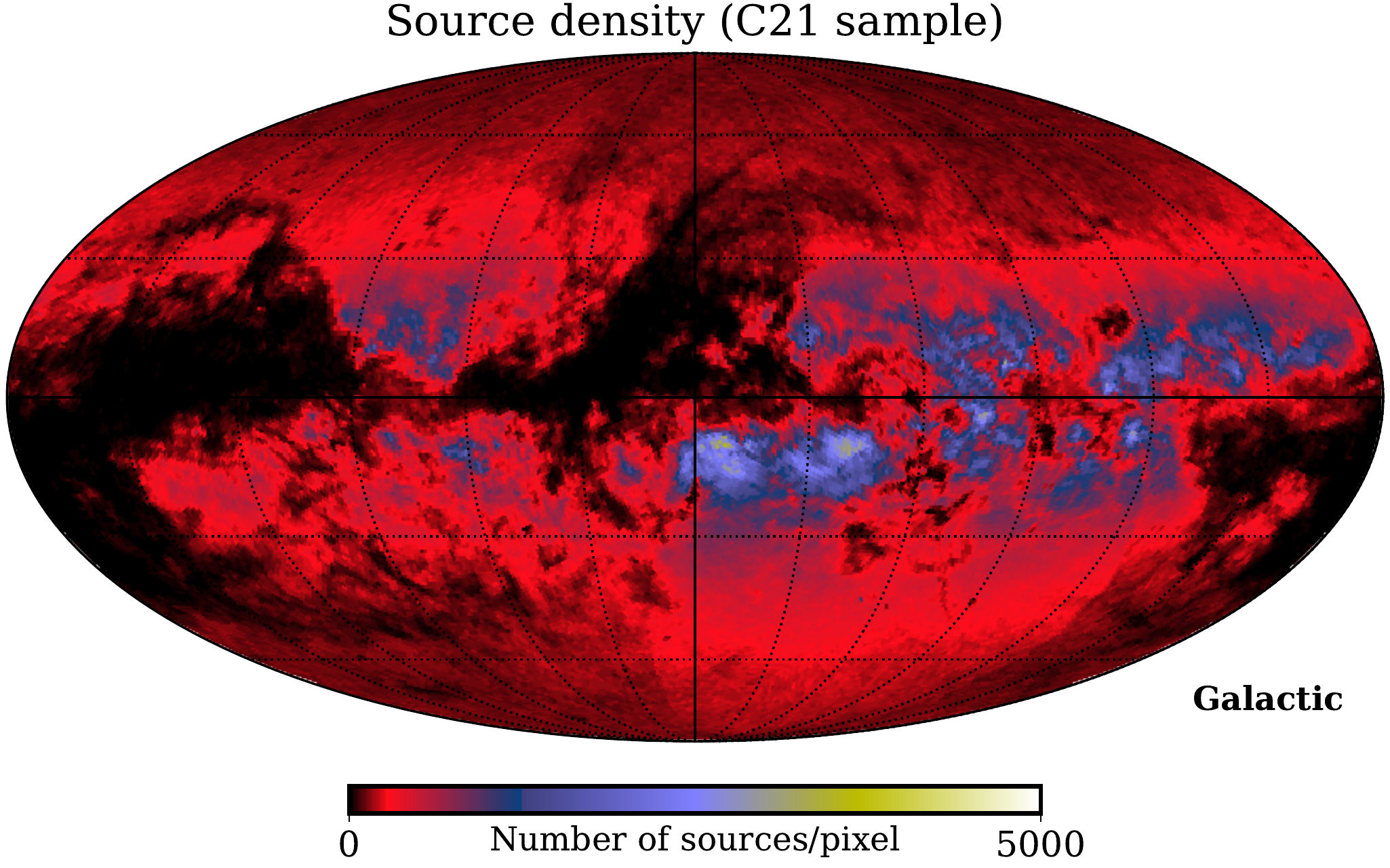}
    
    \bigskip
    
    \includegraphics[width=0.8\textwidth]{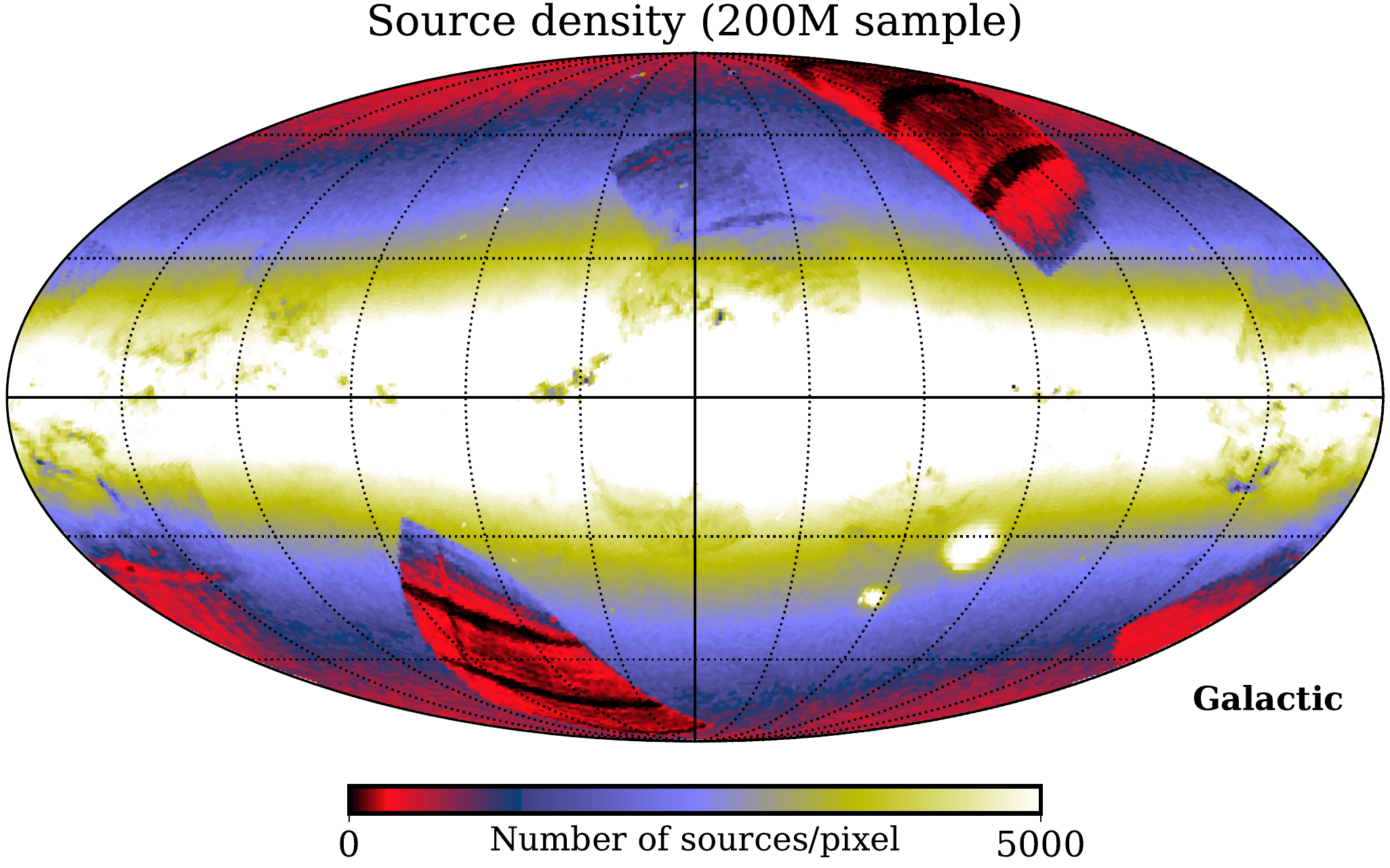}
    \caption{Source density maps, in~Galactic coordinates, corresponding to the C21 (\textbf{top}) and the 200~M (\textbf{bottom}) samples. These maps were created using \textsc{healpix} of level 6 (providing a pixel size of 0.84 square degrees) and~are colour coded depending on the number of sources within the pixel. Note that the colour scale is the same in both maps. See Figure~\ref{fig:mollweide_histogram} for a histogram comparison of the source density in these~maps.}
    \label{fig:mollweide_maps}
\end{figure}

With this work, we move from the $\sim 15$~million sources in C21 to more than 200~million objects. This can be easily visualized in the maps displayed in Figure~\ref{fig:mollweide_maps}, which represent the density of sources on the celestial sphere in Galactic coordinates (using a Mollweide projection with {\sc healpix} of level 6 with~a pixel size of 0.84~square degrees). Not surprisingly, the~200 M sample is concentrated towards the Galactic plane, not being affected by the scarcity of sources in the directions of high interstellar extinction as seen with the C21~sample.

\begin{figure}[H]
    \includegraphics[width=0.8\textwidth]{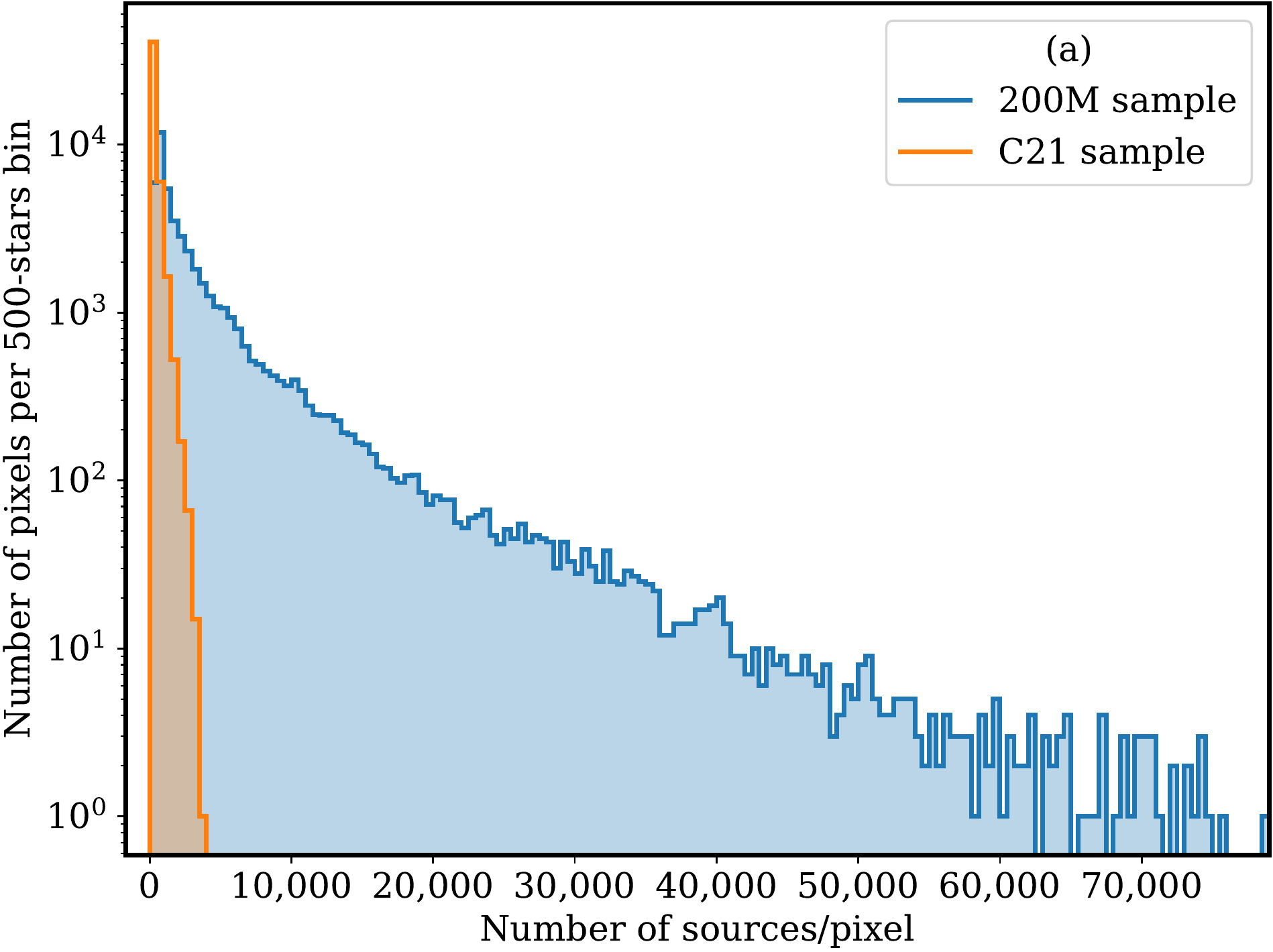}

    \bigskip
    
    \includegraphics[width=0.8\textwidth]{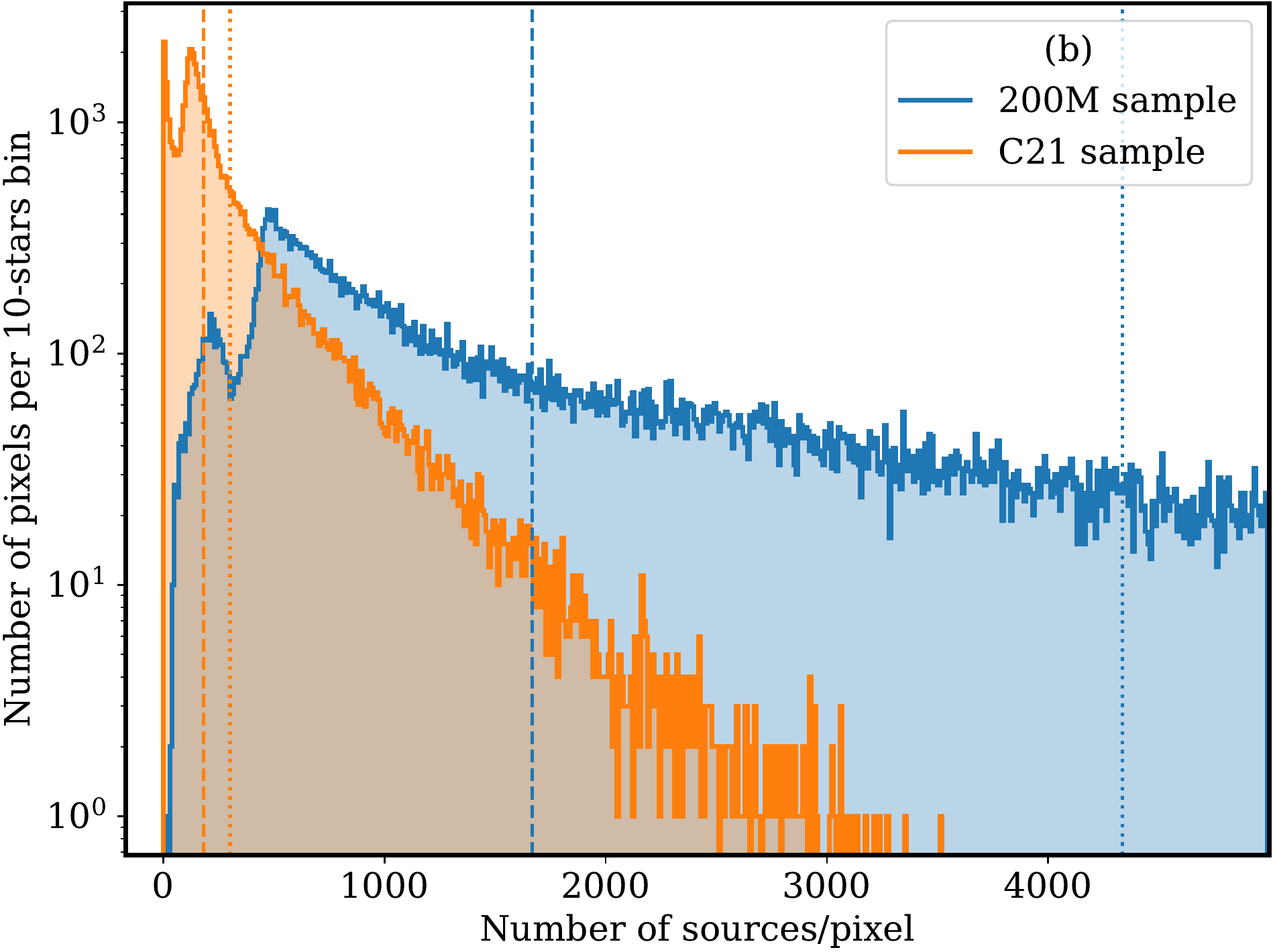}
    
    \caption{Panel~(\textbf{a}): histograms of the source density (i.e., number of sources/pixel, where each pixel corresponds to 0.84 square degrees) in the maps displayed in Figure~\ref{fig:mollweide_maps} for the 200 M sample (blue) and for the C21 sample (orange). Panel~(\textbf{b}): zoom of the previous plot near the origin. The~vertical lines indicate the mean values (302 and 4335 objects/pixel for~C21 and 200 M, respectively; dotted lines) and~median values (183 and 1668 objects/pixel for~C21 and 200 M; dashed lines) for each sample. Note that a different bin size is employed in each panel as~indicated in the label of the vertical~axis.}
    \label{fig:mollweide_histogram}
\end{figure}

The corresponding histograms of the source density are displayed in Figure~\ref{fig:mollweide_histogram}. Panel~(a) shows that, as~expected, there are many directions in the celestial sphere in which the density of stars corresponding to the 200 M sample is clearly larger than as shown by the sources in the C21 sample. The~zoom near the origin, panel~(b), reveals that the density distribution of the C21 sample is bimodal with~a first peak in the first histogram bin, corresponding to the interval $[0,\,10]$ objects/pixel and~a second peak in the interval $[120,\,130]$ objects/pixel. The~first peak corresponds to the directions of high extinction that are purposely excluded by C21. 

The~200 M histogram is also bimodal with~a first peak in the interval $[210,\,240]$ objects/pixel and~a higher second peak in the interval $[470,\,480]$ sources/pixel. This reflects the selection criteria chosen to publish XP spectra for {\gdr3} (see De Angeli et al.~\cite{DeAngeli2022}). In both samples, the histogram distributions are clearly asymmetric  as~indicated by the location of their respective means (vertical dotted lines) and medians (vertical dashed lines), whose values are given in the figure caption. 

Considering that the pixel size employed in the maps displayed in Figure~\ref{fig:mollweide_maps} is slightly
below \mbox{1 square} degree, we confirm that the 200~M sample offers several hundred sources per square~degree in most regions of the sky. A~more detailed calculation indicates that  99.7\% of the celestial sphere is covered with a source density above 100 sources/(square degree) by this catalogue. The rest of the sky with less sources present corresponds to regions in the {\gaia} mission with less transits and without {\gaia} spectrophotometry available in {\gdr3}~\citep{DeAngeli2022}.

\begin{figure}[H]
    \includegraphics[width=0.8\textwidth]{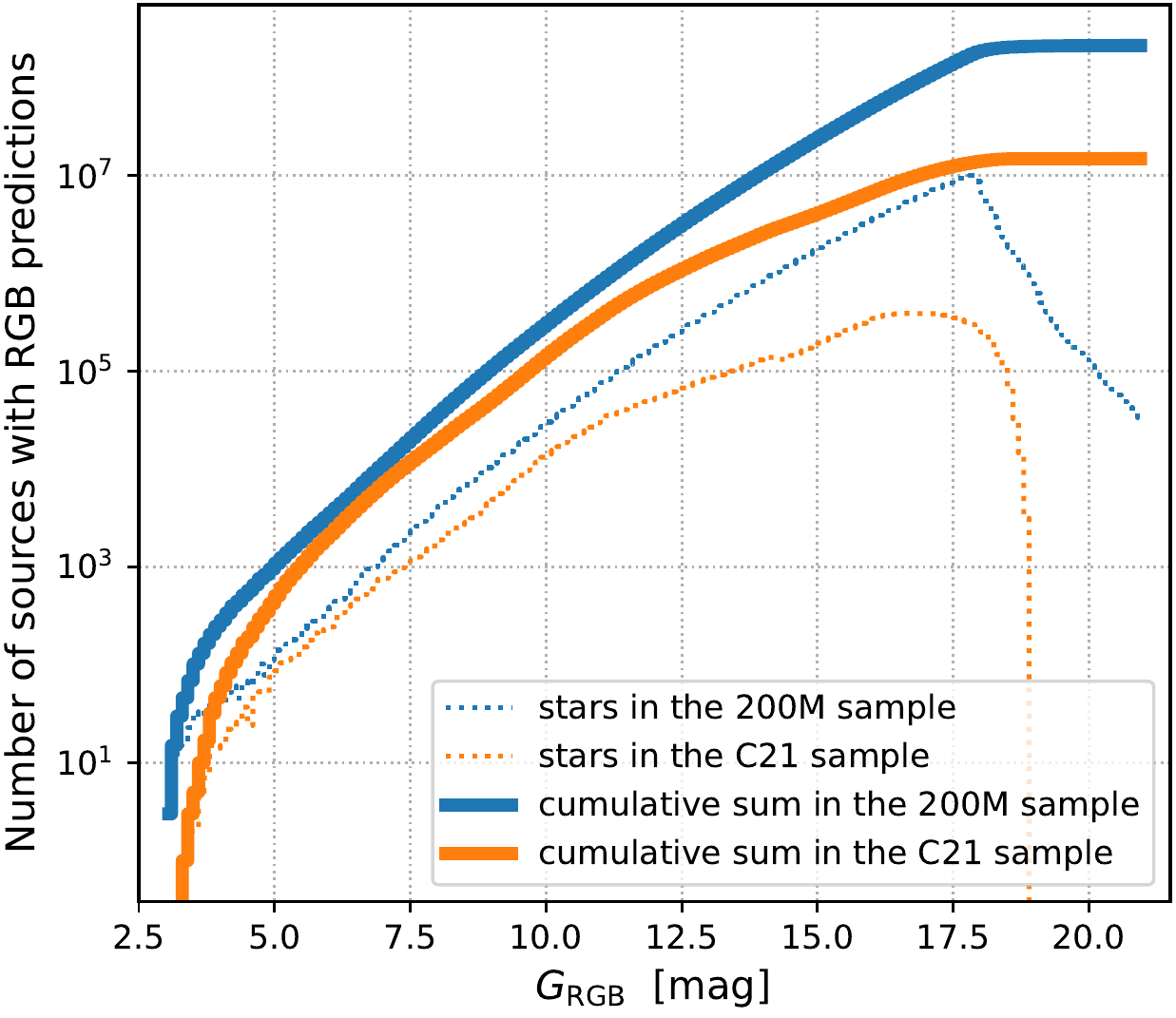}
    \caption{Variation in the number of sources with RGB magnitude predictions as a function of $G_{\rm RGB}$ for~both the 200 M (blue) and the C21 (orange) samples. The~dotted lines show the histograms in bins of 0.1~mag, whereas the thick full lines display the cumulative~sums.}
    \label{fig:hist_G_cumsum_200M_C21}
\end{figure}

Clearly, the~above numbers should be taken with some caution because, in a practical way, the number of usable sources will also be a function of the magnitude limit reached. In~this sense, Figure~\ref{fig:hist_G_cumsum_200M_C21} displays the variation in the number of sources as a function of the magnitude in the $G_{\rm RGB}$ band. The~plot shows the histogram (dotted lines) and the cumulative sum (thick full lines) for~the C21 (orange) and the 200 M samples (blue). In~both samples, there is a sudden decrease in the number of sources for $G_{\rm RGB} \gtrsim 17.5$~mag. 

This is because most of the sources published in {\gdr3} were selected to have $G_{\rm Gaia}<17.65$~mag, although~the XP spectra for some sources (including white dwarfs, galaxies and quasars) were also explicitly published above this magnitude limit  (see De Angeli et al.~\cite{DeAngeli2022}). The~cumulative number of stars down to some particular $G_{\rm RGB}$ values are listed in \mbox{Table~\ref{tab:sums}}. The~M200 sample clearly outnumbers the C21 sample in number of sources at any magnitude~value.

\unskip

\begin{table}[H]
\setlength{\tabcolsep}{9.9mm}
    \caption{Cumulative sums down to some particular $G_{\rm RGB}$ value for the C21 and the 200 M~samples.}
    \label{tab:sums}
 \begin{tabularx}{\textwidth}{crr}
 \toprule
 \boldmath{\textbf{$G_{\rm RGB}$ Interval [mag]}}& 
\multicolumn{1}{c}{\textbf{Sources in C21}} & 
\multicolumn{1}{c}{\textbf{Sources in 200 M}}\\
\midrule
 <10 &    133,019\;\;\; &     299,041\;\;\;\\
 <11 &    344,349\;\;\; &     775,526\;\;\;\\
 <12 &    761,279\;\;\; &   1,948,176\;\;\;\\
 <13 &  1,431,314\;\;\; &   4,636,535\;\;\;\\
 <14 &  2,484,448\;\;\; &  10,495,506\;\;\;\\
 <15 &  3,977,325\;\;\; &  22,967,532\;\;\;\\
 <16 &  6,546,925\;\;\; &  48,355,681\;\;\;\\
 <17 & 10,346,030\;\;\; &  97,464,183\;\;\;\\
 <18 & 13,820,929\;\;\; & 182,081,668\;\;\;\\
 <19 & 14,854,959\;\;\; & 209,094,405\;\;\;\\
 <20 & 14,854,959\;\;\; & 212,129,751\;\;\;\\
$\lesssim$21 & 14,854,959\;\;\; & 213,064,002\;\;\;\\
\bottomrule
    \end{tabularx}

\end{table}
\unskip

\subsection{Source~Characteristics}

\begin{figure}[H]
    \includegraphics[width=0.8\textwidth]{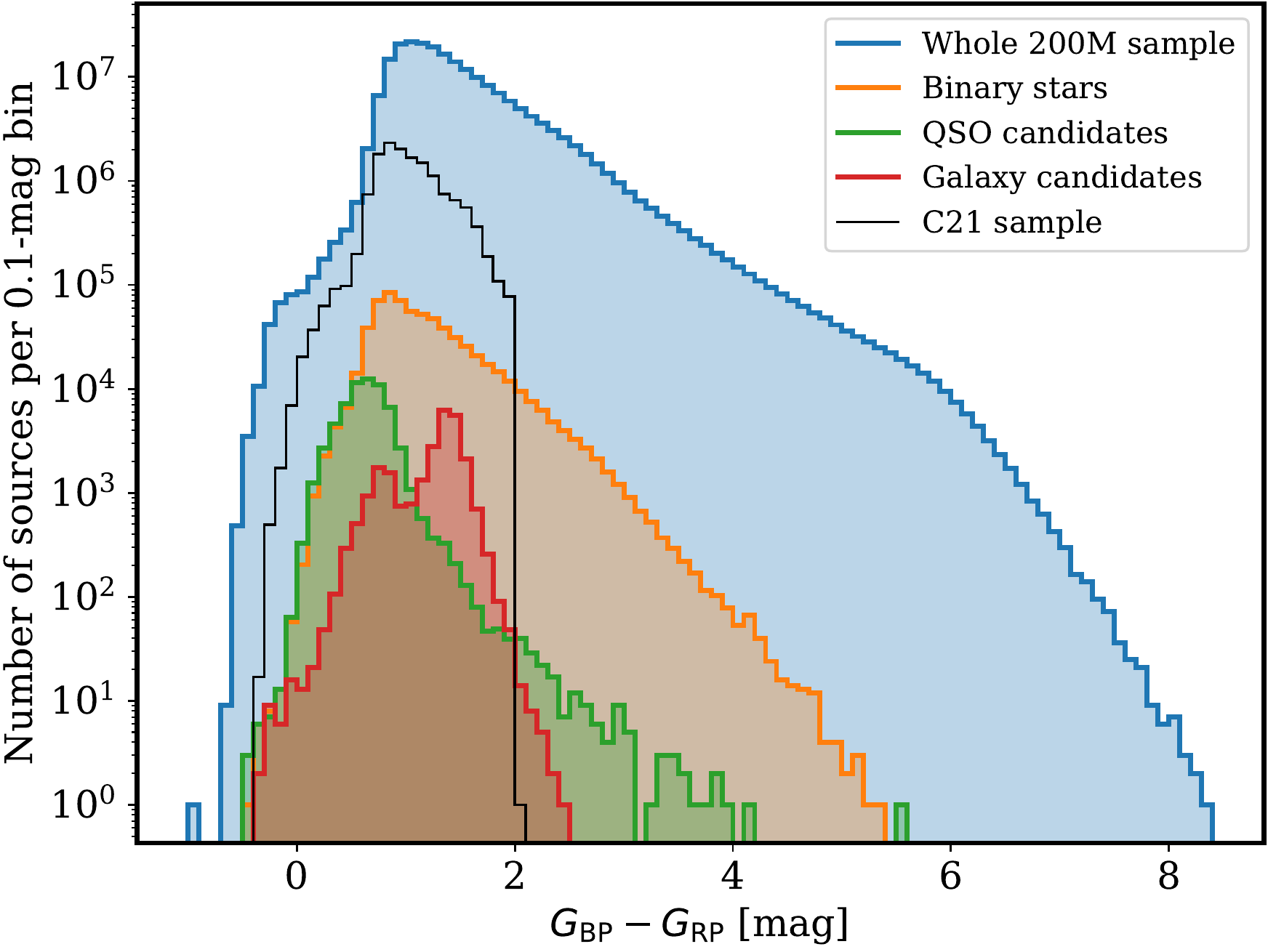}
    \caption{Histogram of the 200 M sample as a function of \mbox{${\rm G}_{\rm BP}-{\rm G}_{\rm RP}$} colour in bins of 0.1~mag. The~whole 200 M sample is displayed in blue, whereas the subsamples corresponding to sources flagged in {\gdr3} as {\texttt{non\_single\_star}, \texttt{in\_qso\_candidates} and~\texttt{in\_galaxy\_candidates}} are displayed in orange, green and red, respectively. In~addition, the~histogram corresponding to the C21 sample is displayed with a thin black line, which is limited to \mbox{$-0.5 \leq {\rm G}_{\rm BP}-{\rm G}_{\rm RP} \leq 2.0$~mag}.}
    \label{fig:hist_bp_rp}
\end{figure}

\textls[-15]{Apart from the different number of sources included in the 200 M and C21 samples, important differences in some specific characteristics of the sources are worth mentioning. In~particular, the~C21 subsample was restricted to sources with \mbox{$-0.5 \leq {\rm G}_{\rm BP}-{\rm G}_{\rm RP} \leq 2.0\,{\rm mag}$}}, whereas the 200 M sample does not include this constraint. This is clearly manifested in the histogram displayed in Figure~\ref{fig:hist_bp_rp}, which shows that the new 200 M sample (blue filled histogram) also includes much redder objects that were not included in the C21 sample (black line). 

In~addition, this figure also represents the histograms corresponding to the 200 M subsamples corresponding to particular sources not classified as simple stars in {\gdr3} (namely the sources flagged as {\texttt{non\_single\_star}, \texttt{in\_qso\_candidates} and~\texttt{in\_galaxy\_candidates}}), for~which the polynomial RGB calibration in C21, derived only for single stars, is not~suitable.

\begin{figure}[H]
    \centering
    \includegraphics[width=0.45\textwidth]{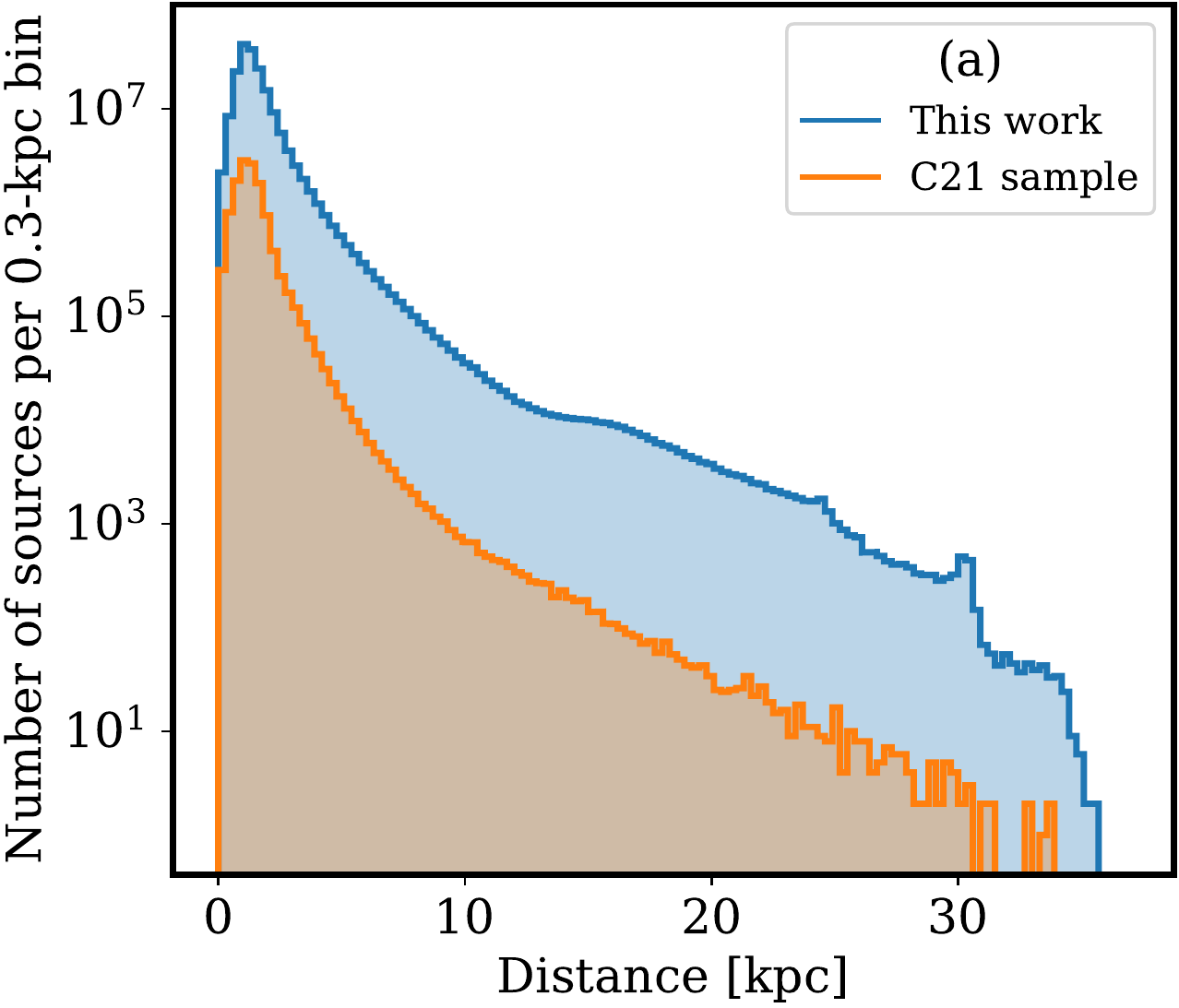}

    \bigskip
    
    \includegraphics[width=0.45\textwidth]{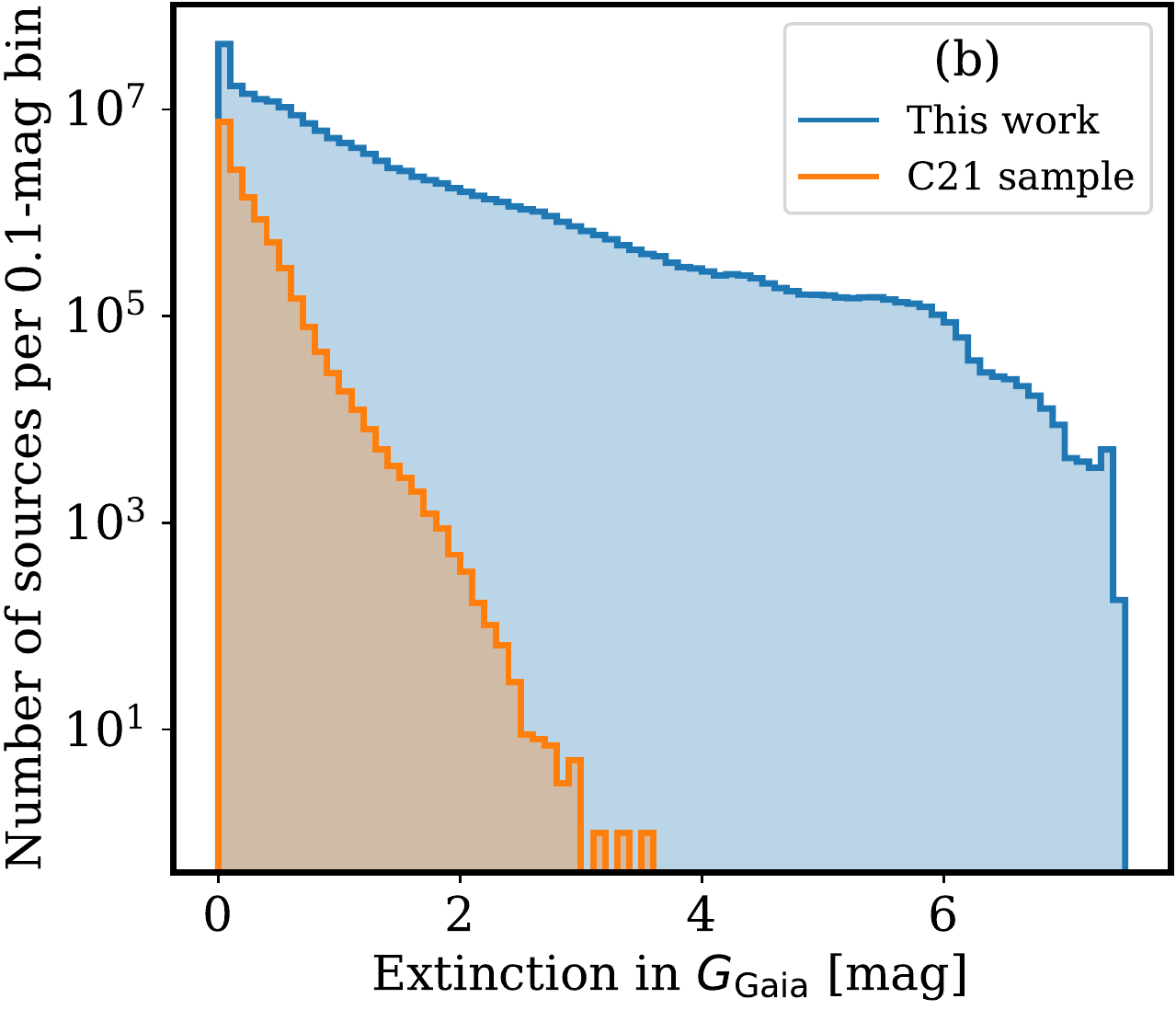}
    \hfill
    \includegraphics[width=0.45\textwidth]{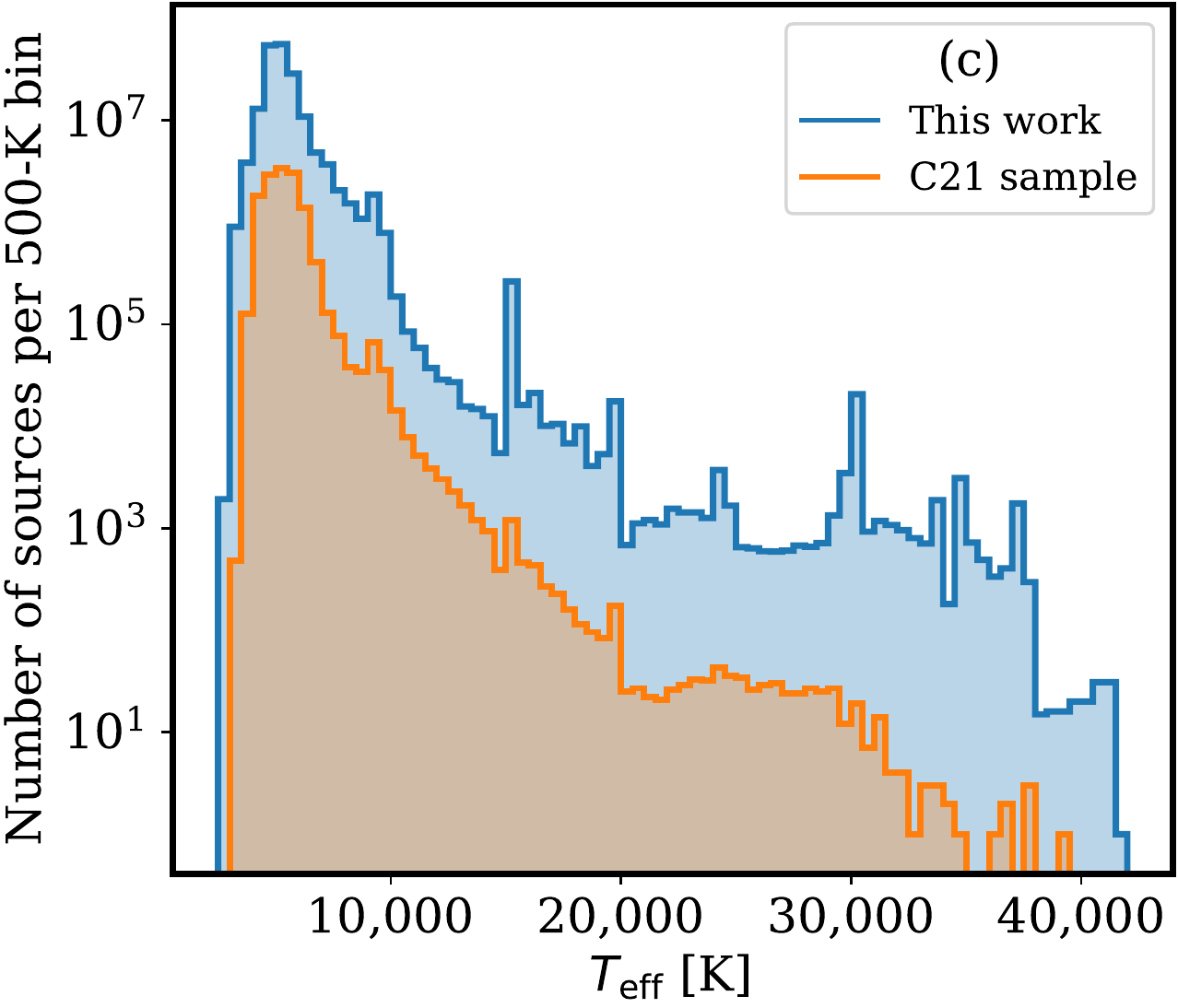}

    \bigskip
    
    \includegraphics[width=0.45\textwidth]{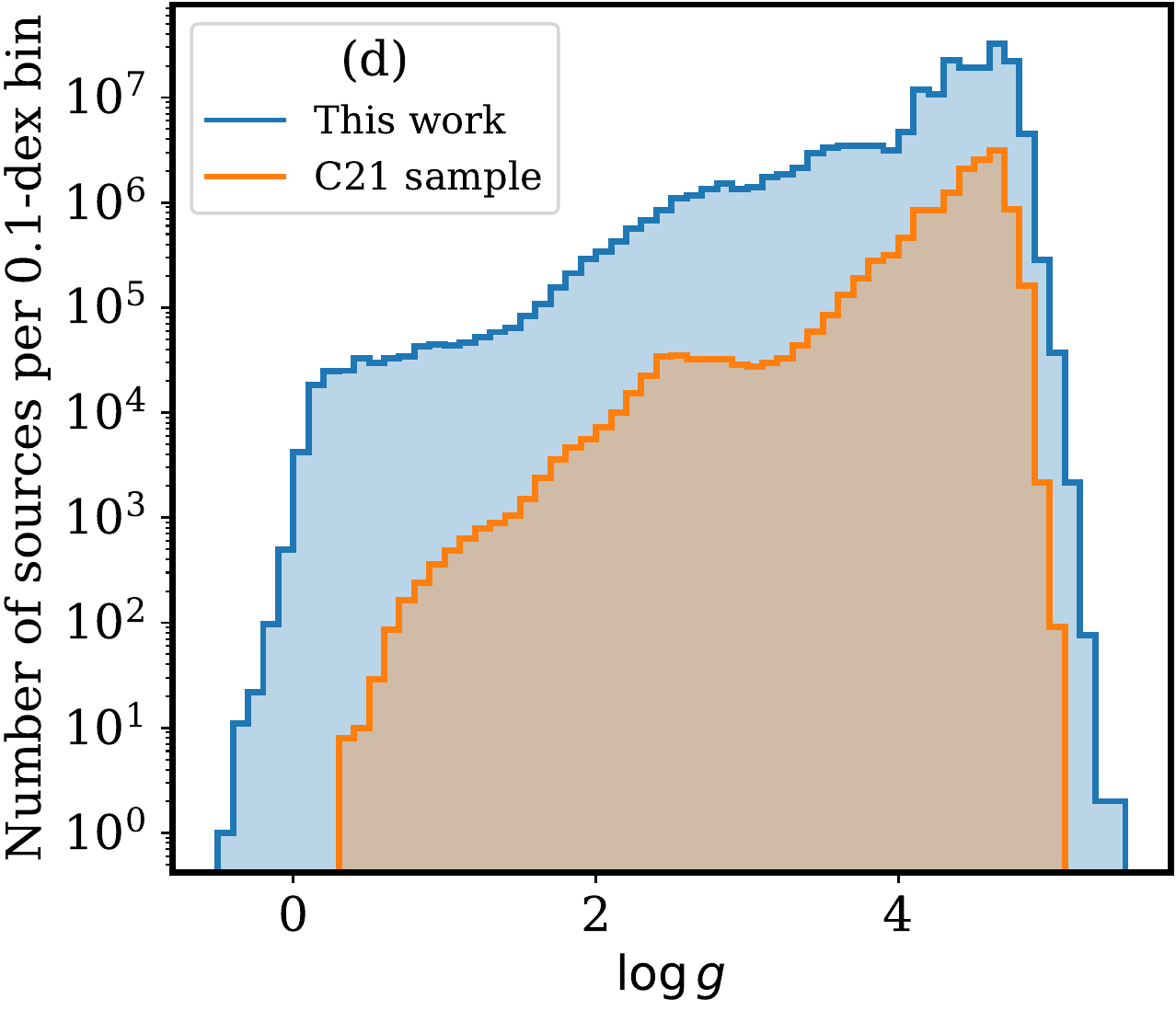}
    \hfill
    \includegraphics[width=0.45\textwidth]{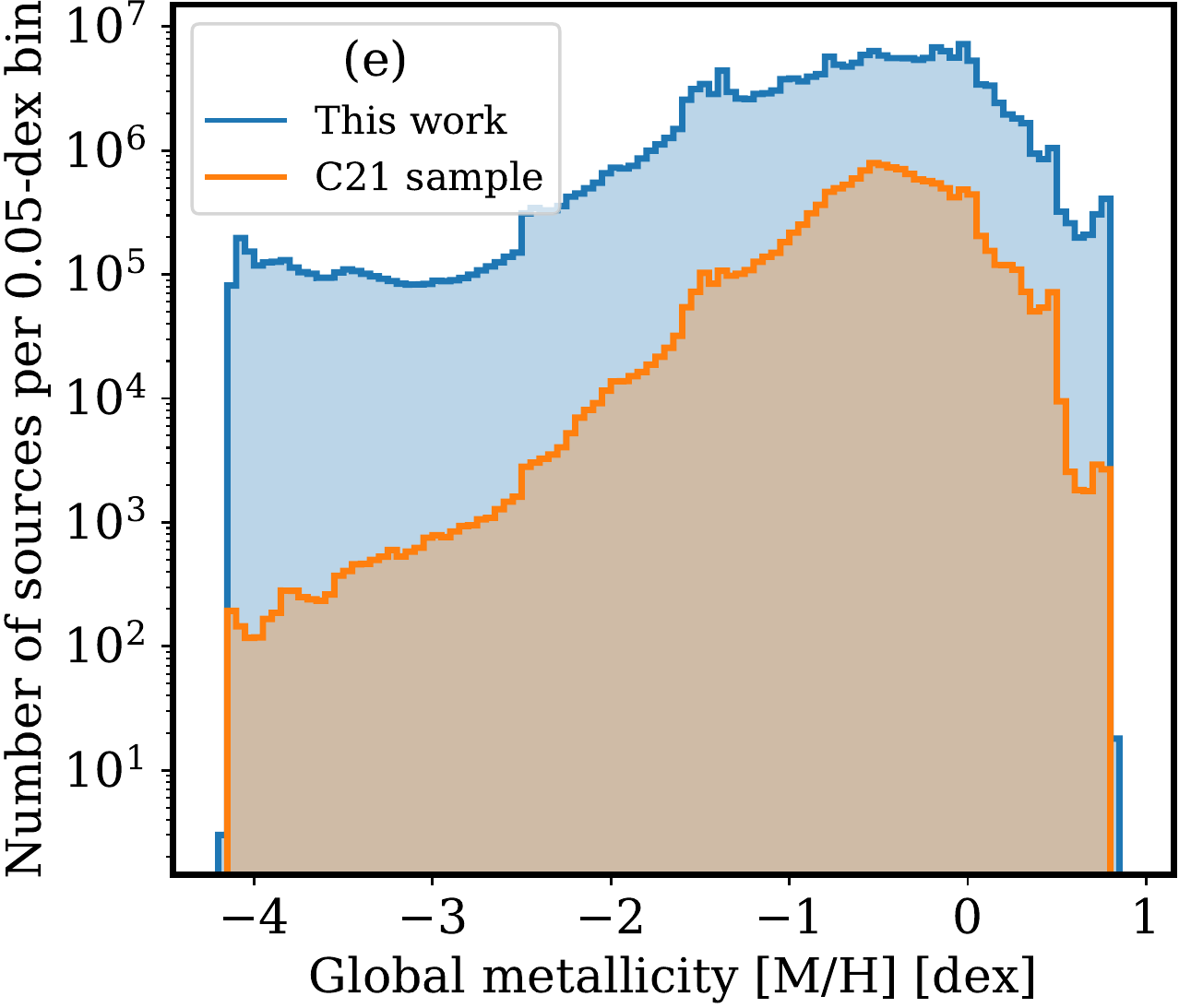}
    
    \caption{Comparison between the 200 M sample built in this work (blue histograms) and the C21 sample (orange histograms). The~panels represent different parameters retrieved from the {\gdr3} database, namely \texttt{distance\_gspphot} (in kpc, panel~(\textbf{a})) and extinction in the ${\rm G}_{\rm Gaia}$ band \texttt{ag\_gspphot} (in magnitudes, panel~(\textbf{b})) as well as~typical stellar parameters, such as the effective temperature \texttt{teff\_gspphot} (in K, panel~(\textbf{c})), surface gravity \texttt{logg\_gspphot} (logarithm of cgs units, panel~(\textbf{d})) and~global metallicity \texttt{mh\_gspphot} (dex units, panel~(\textbf{e})).}
    \label{fig:histograms_200M_C21}
\end{figure}

We also explore, in Figure~\ref{fig:histograms_200M_C21}, the histogram distribution of some additional parameters derived from the {\gdr3} data, in~particular, the {\texttt{distance\_gspphot}} (in kpc, panel~(a)) and extinction in the ${\rm G}_{\rm Gaia}$ band {\texttt{ag\_gspphot}} (in magnitudes, panel~(b)) as well as~typical stellar parameters, such as the effective temperature {\texttt{teff\_gspphot}} (in K, panel~(c)), surface gravity {\texttt{logg\_gspphot}} (logarithm of cgs units, panel~(d)) and~global metallicity {\texttt{mh\_gspphot}} (dex units, panel~(e)). 
The 200 M sample contains proportionally more distant sources than does the C21 sample, although~the most noticeable difference is the inclusion of more sources with much larger extinction. The~200~M catalogue incorporates proportionally more stars with large effective temperature, low surface gravity and~low~metallicity.

\unskip

\subsection{Magnitude~Residuals}
\label{sec:residuals}

\begin{figure}[H]
    \includegraphics[width=0.48\textwidth]{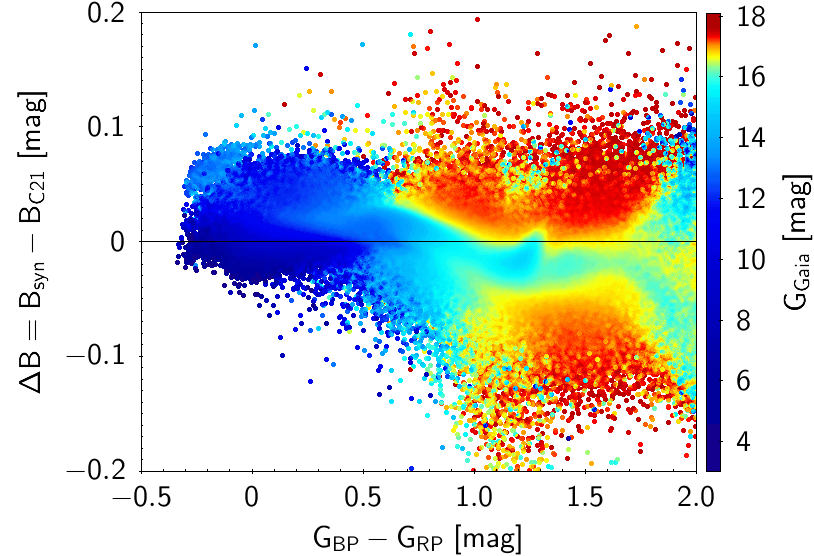}
    \hfill
    \includegraphics[width=0.48\textwidth]{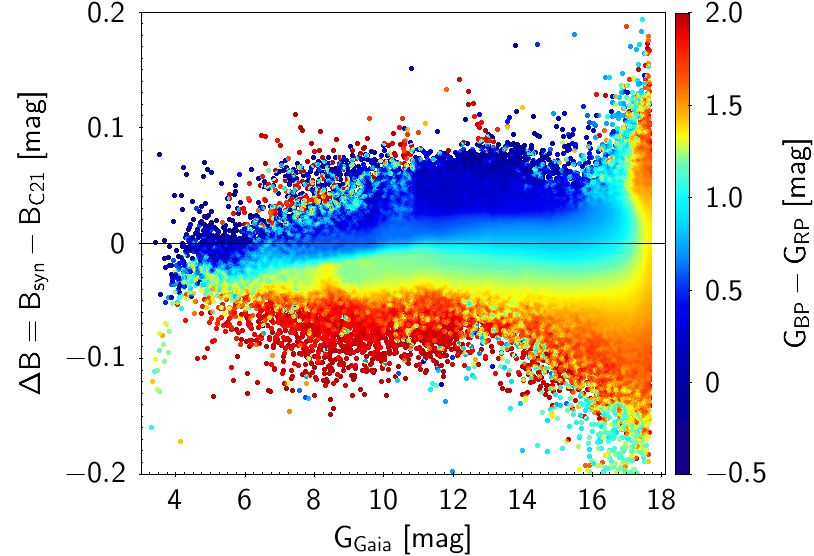}

    \bigskip
    
    \includegraphics[width=0.48\textwidth]{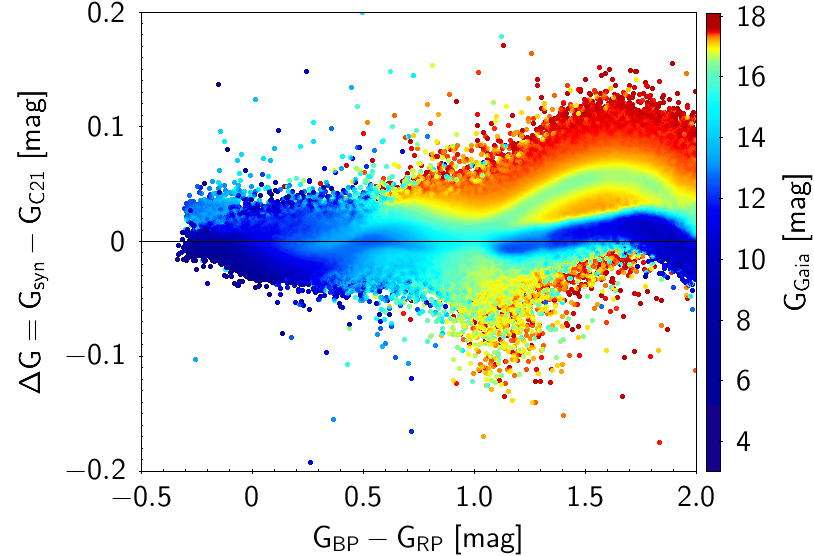}
    \includegraphics[width=0.48\textwidth]{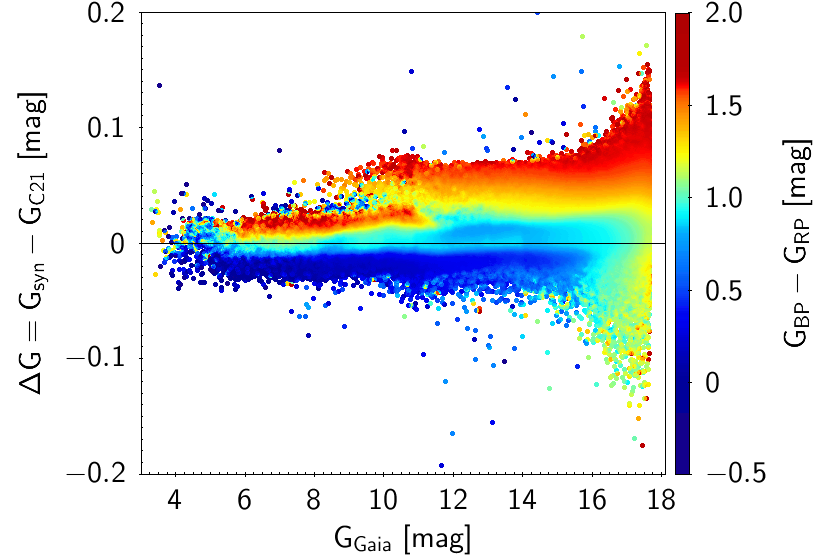}

    \bigskip
    
    \includegraphics[width=0.48\textwidth]{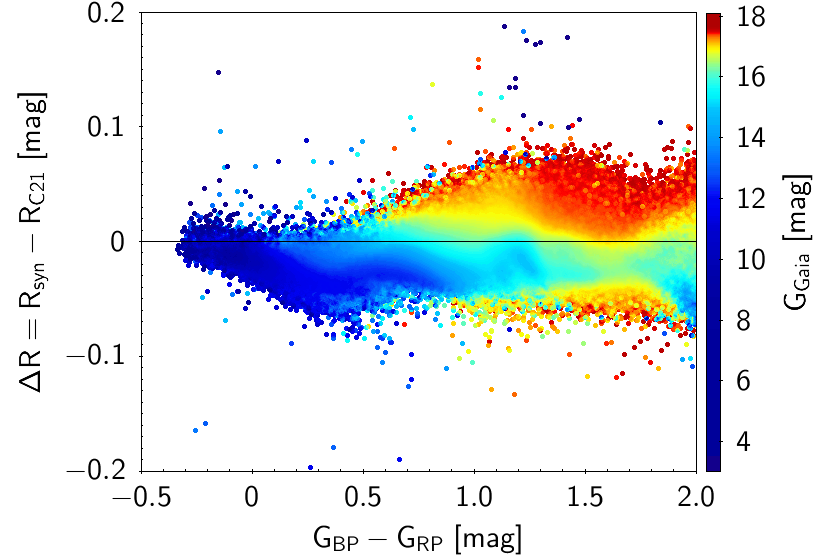}
    \includegraphics[width=0.48\textwidth]{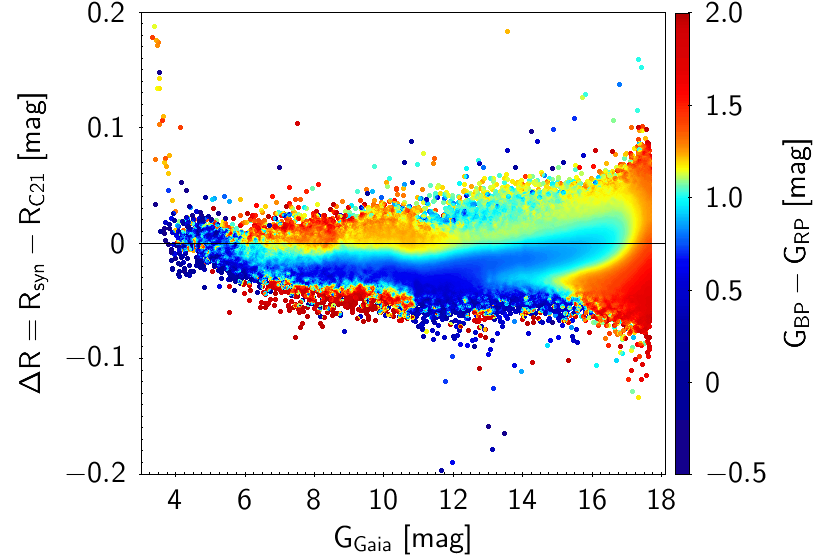}
    
    \caption{Difference between the synthetic magnitude computed in this work ($X_{\rm syn}$) and those provided by C21 ($X_{C21}$) for $X=B_{\rm RGB}$ (\textbf{top row}), $G_{\rm RGB}$ (\textbf{middle row}) and $R_{\rm RGB}$ (\textbf{bottom row}) as a function of {\gaia} $G_{\rm BP}-G_{\rm RP}$ colour (\textbf{left column}) and $G_{\rm Gaia}$ magnitude (\textbf{right column}) for non-variable sources in~common.}
    \label{fig:delta}
\end{figure}

\begin{figure}[H]
    \includegraphics[width=0.48\textwidth]{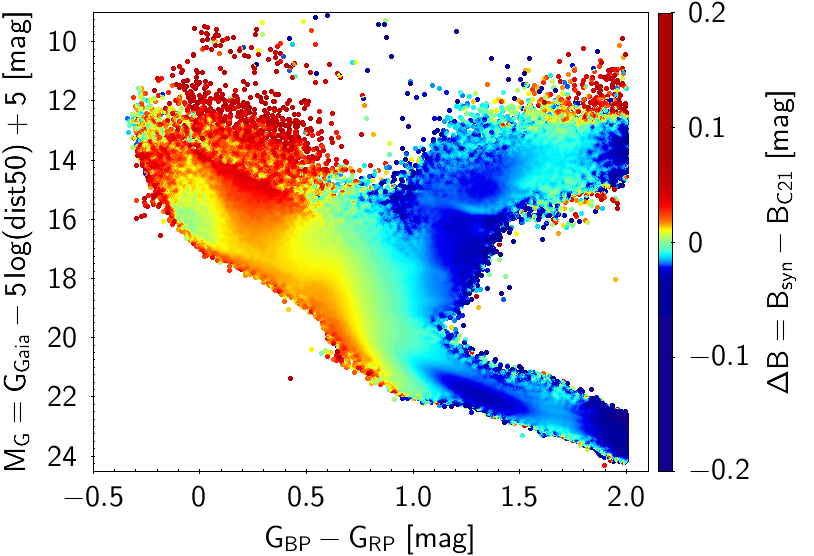}
    \hfill
    \includegraphics[width=0.48\textwidth]{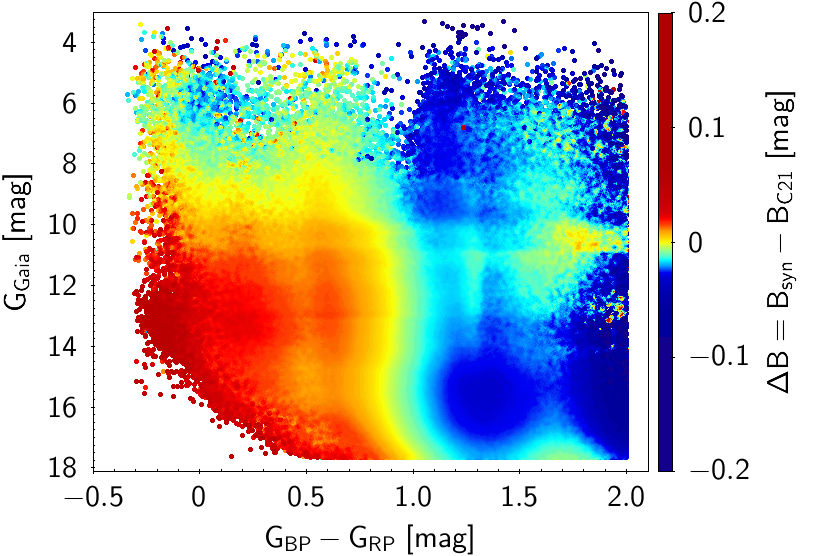}

    \bigskip

    \includegraphics[width=0.48\textwidth]{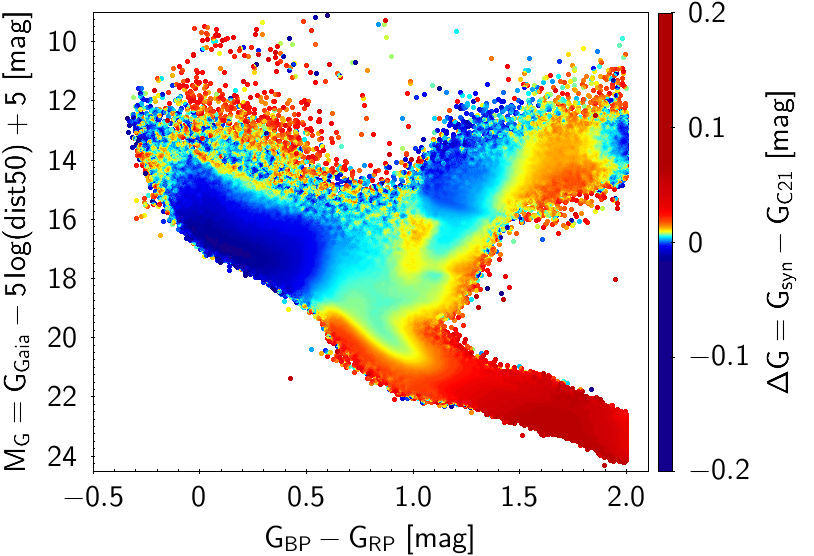}
    \hfill
    \includegraphics[width=0.48\textwidth]{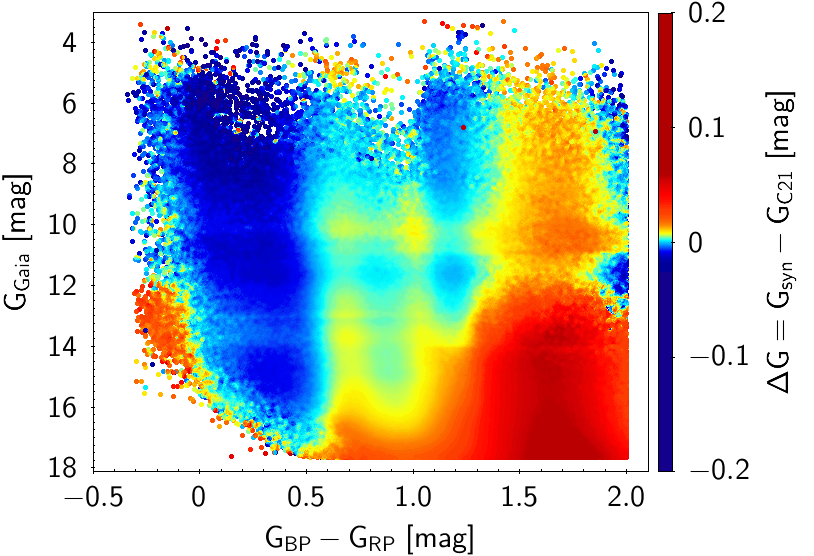}

    \bigskip
    
    \includegraphics[width=0.48\textwidth]{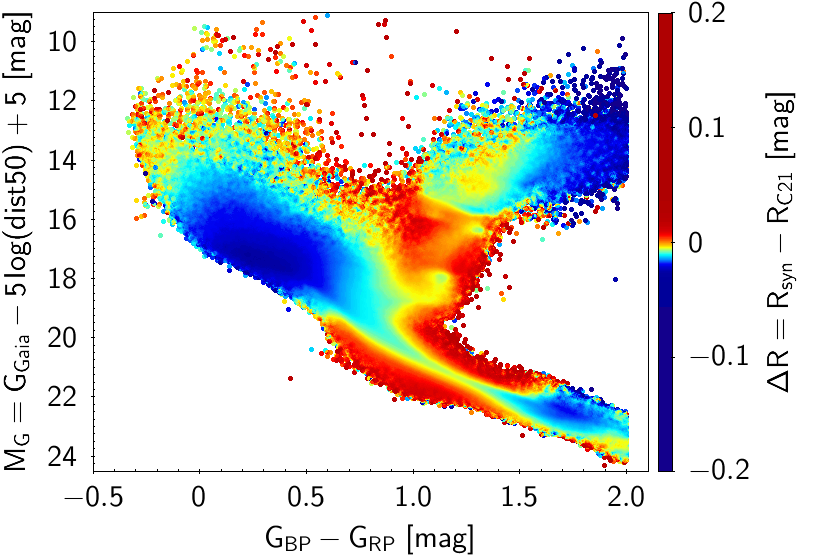}
    \hfill
    \includegraphics[width=0.48\textwidth]{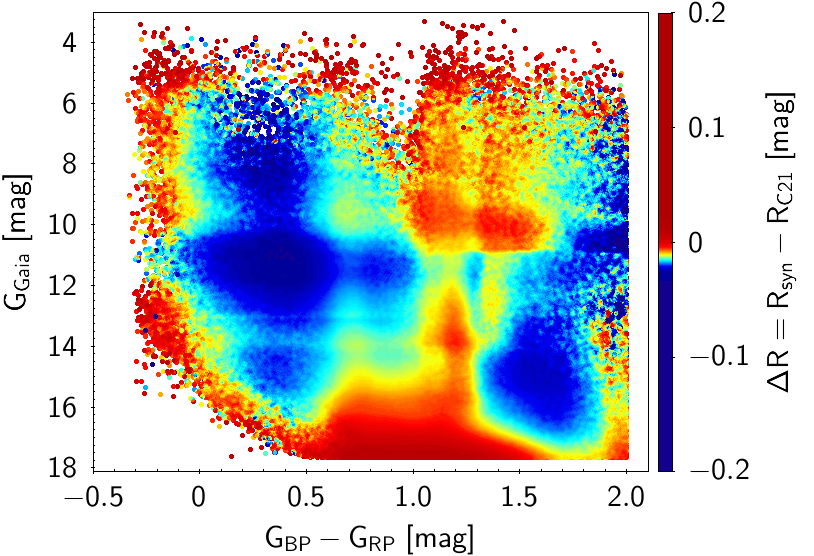}
    
    \caption{Absolute (\textbf{left column}) and apparent (\textbf{right column}) {\gaia} magnitude-colour diagrams for non-variable sources in common with C21. The~colour index shows the difference between the synthetic magnitude ($X_{\rm syn}$) and those provided by C21 ($X_{\rm C21}$) for $X=B_{\rm RGB}$ (\textbf{top row}), $G_{\rm RGB}$ (\textbf{middle row}) and $R_{\rm RGB}$ (\textbf{bottom row}).}
    \label{fig:deltaHR}
\end{figure}

As the synthetic photometry in the 200 M sample is derived directly from the observed spectra and not predicted based on broad band photometry, the~values obtained in this work should be more accurate than those provided by C21. Thus, the~level of discrepancy should be mainly constrained by the precision of C21. We compare, in Figures~\ref{fig:delta} and \ref{fig:deltaHR}, the synthetic photometry derived here with the predictions provided by C21 for \mbox{12,920,293} non-variable
sources in common. As~expected, the~discrepancies mostly fall in the $\pm 0.1$~mag range, which was the required accuracy claimed in C21. This provides confidence in the validity of our~results.

\unskip

The observed biases and colour trends in the differences shown in \mbox{Figures~\ref{fig:delta} and \ref{fig:deltaHR}} are due to systematics in the {\gaia} spectra (see Gaia Collaboration et al.~\cite{Montegriffo2022SynthPhot}) and the photometric transformations proposed by C21---being more important for the latter.
{{The `bridge' structure} present in the $G_{\rm RGB}$ residual (middle-left panel in Figure~\ref{fig:delta}) with $1.0<G_{\rm BP}-G_{\rm RP}<2.0$~mag  contains main sequence stars, although~giants have lower levels of residuals).} 

On the other hand, the results in the right panels of Figures~\ref{fig:delta} and \ref{fig:deltaHR} show the different behaviour present for $G_{\rm Gaia}<11.5$~mag with respect to fainter magnitudes. This feature is likely induced by {\gaia} XP data  as~sources with $G_{\rm Gaia}<11.5$~mag are observed with different gating strategies in {\gaia} to avoid saturation {{(see}~\cite{Prusti2016,Crowley2016})}. For~these bright {\gaia} sources, larger systematics are expected in the XP spectra  due to the lower number of calibrators present for those special gating conditions {{used to minimise} saturation events in {\gaia}}.

\begin{figure}[H]
    \includegraphics[width=0.55\textwidth]{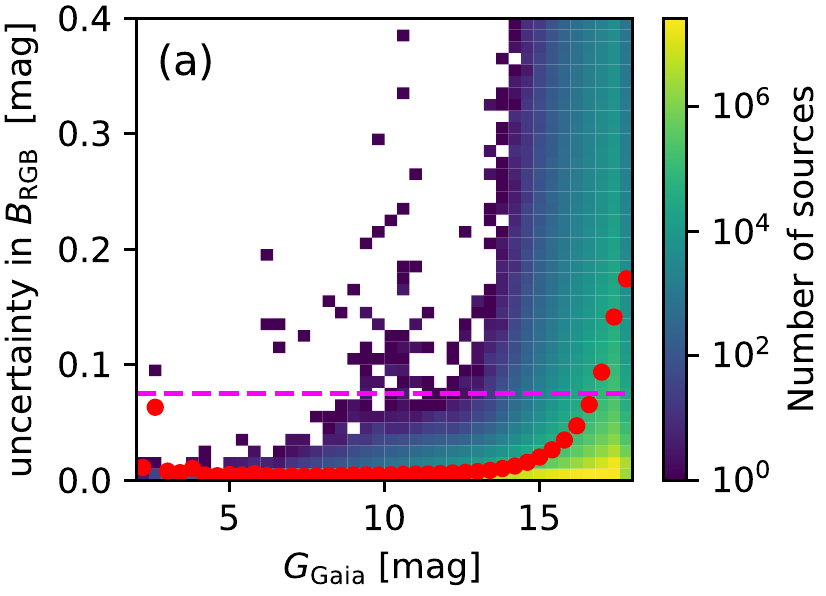}

    \bigskip
    
    \includegraphics[width=0.55\textwidth]{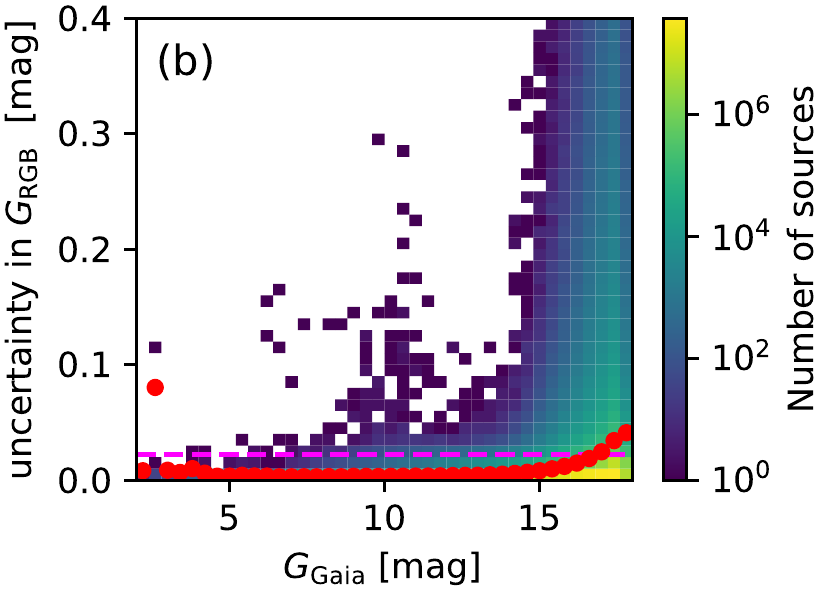}

    \bigskip 
    
    \includegraphics[width=0.55\textwidth]{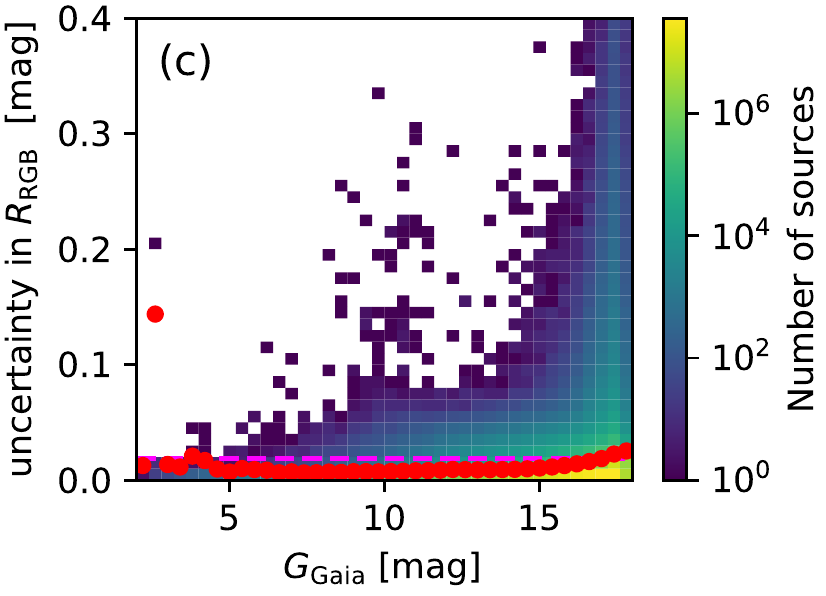}
    \caption{2D histograms showing the number of sources as a function of $G_{\rm Gaia}$ (the horizontal axis; using 0.4~mag bins) and the uncertainties in the synthetic $B_{\rm RGB}$ (panel~(\textbf{a})), $G_{\rm RGB}$ (panel~(\textbf{b})) and $R_{\rm RGB}$ (panel~(\textbf{c})) magnitudes (vertical axis; using 0.01~mag bins), computed as explained in Section~\ref{sec:methodology}. The~red filled circles indicate the 99th percentile at each bin in the horizontal axis, which reveals that, for most $G_{\rm Gaia}$ bins, these numbers are below 0.01~mag in the vertical axis with~the exception of one bin at the bright extreme, \mbox{$2.4 \leq G_{\rm Gaia}\leq 2.8$ mag} and~for the fainter sources beyond \mbox{$G_{\rm Gaia}\gtrapprox 14$ mag}. The~magenta dashed horizontal line indicates the {{99th}} percentile for the whole sample with values of 0.075, 0.022 and 0.019~mag for~$B_{\rm RGB}$, $G_{\rm RGB}$ and $R_{\rm RGB}$, respectively.} 
    \label{fig:density_uncertainties}
\end{figure}

The method for deriving the synthetic RGB magnitudes explained in Section~\ref{sec:methodology} allows for the derivation of the associated uncertainties, and it is possible to examine their behaviour as a function of the relevant parameters. In~particular, Figure~\ref{fig:density_uncertainties} represents 2D histograms with the density of sources as a function of $G_{\rm Gaia}$ (the horizontal axis) and the uncertainties in the synthetic magnitudes $B_{\rm RGB}$, $G_{\rm RGB}$ and $R_{\rm RGB}$ (panels~(a), (b) and~(c), respectively). Not surprisingly, the~uncertainties increase when moving to fainter objects. The~red filled circles mark the 99th percentile at each 0.4~mag bin in the horizontal axis. Interestingly, these numbers are below 0.01~mag for a wide $G_{\rm Gaia}$ interval, increasing beyond $G_{\rm Gaia}\gtrsim 14$~mag.

Since the number of faint sources is larger than the number of bright ones, the~{{99th}} percentiles for the whole sample with values of 0.075, 0.022 and 0.019~mag for $B_{\rm RGB}$, $G_{\rm RGB}$ and $G_{\rm RGB}$, respectively, (represented with the dashed horizontal magenta line in each panel) are naturally larger than the values for bright sources (with lower uncertainties but less abundant in the total sample).
It is important to highlight that, although the uncertainties are unavoidably larger for fainter objects, the~also larger number of available sources at those magnitude regimes should allow the simultaneous observation of many more of them to be used as standards. The~statistical combination of these measurements should help to diminish the uncertainties at the faint~end.

\section{Validity of the C21 Polynomial~Calibration}
\label{sec:validity_C21_polynomial_calibration}

\begin{figure}[H]
    \includegraphics[width=0.65\textwidth]{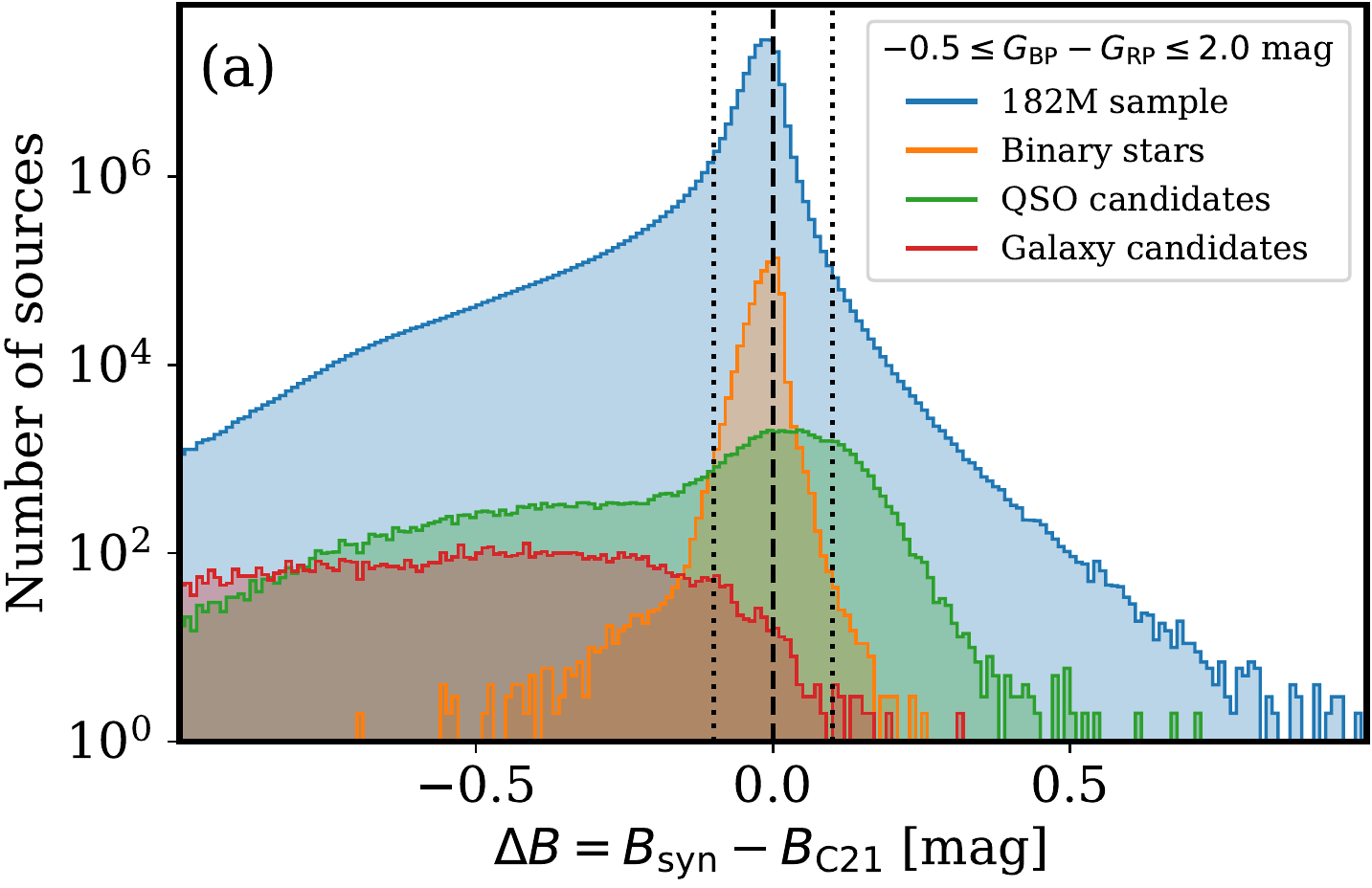}
    
    \bigskip
    
    \includegraphics[width=0.65\textwidth]{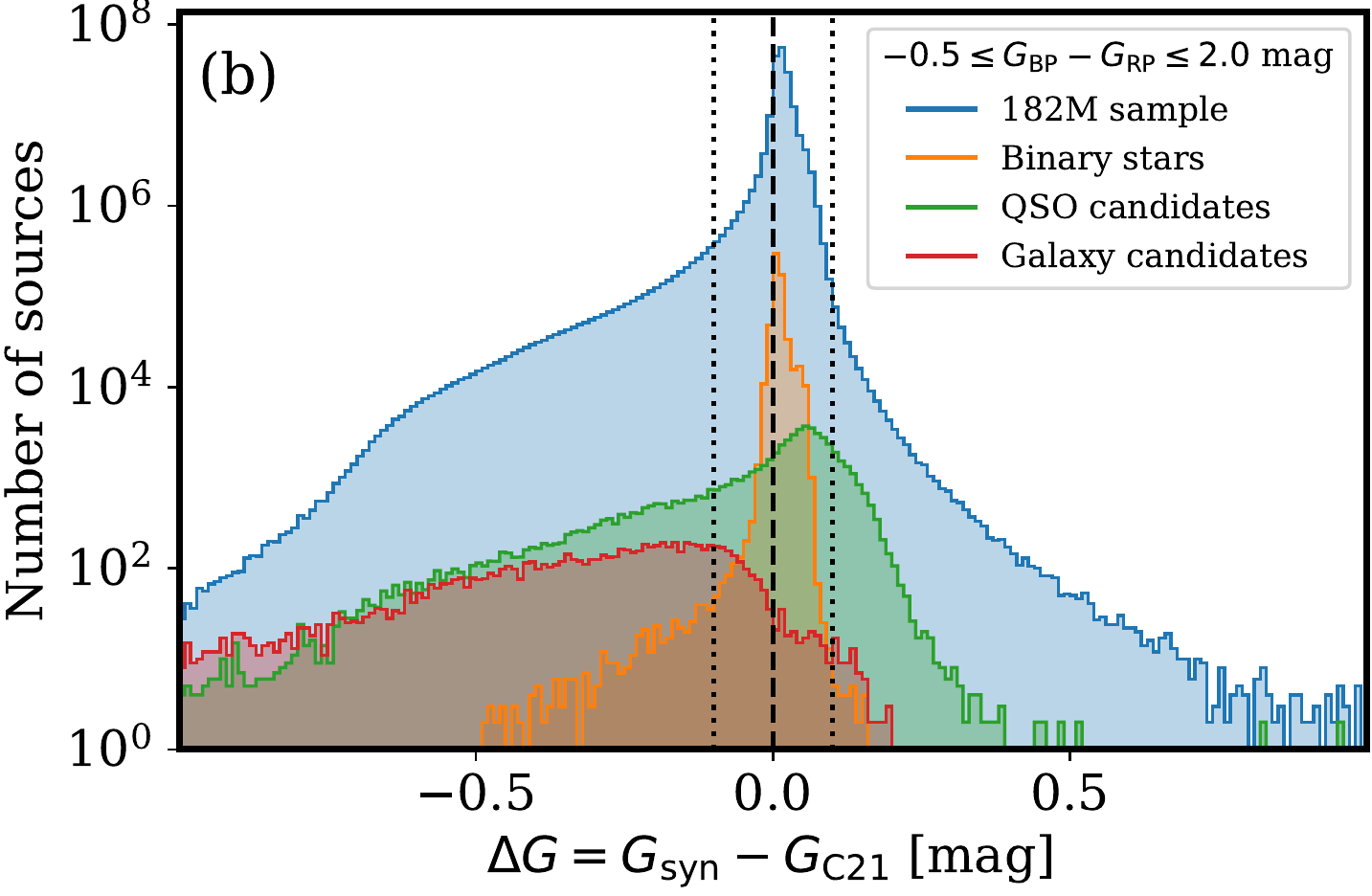}
    
    \bigskip
    
    \includegraphics[width=0.65\textwidth]{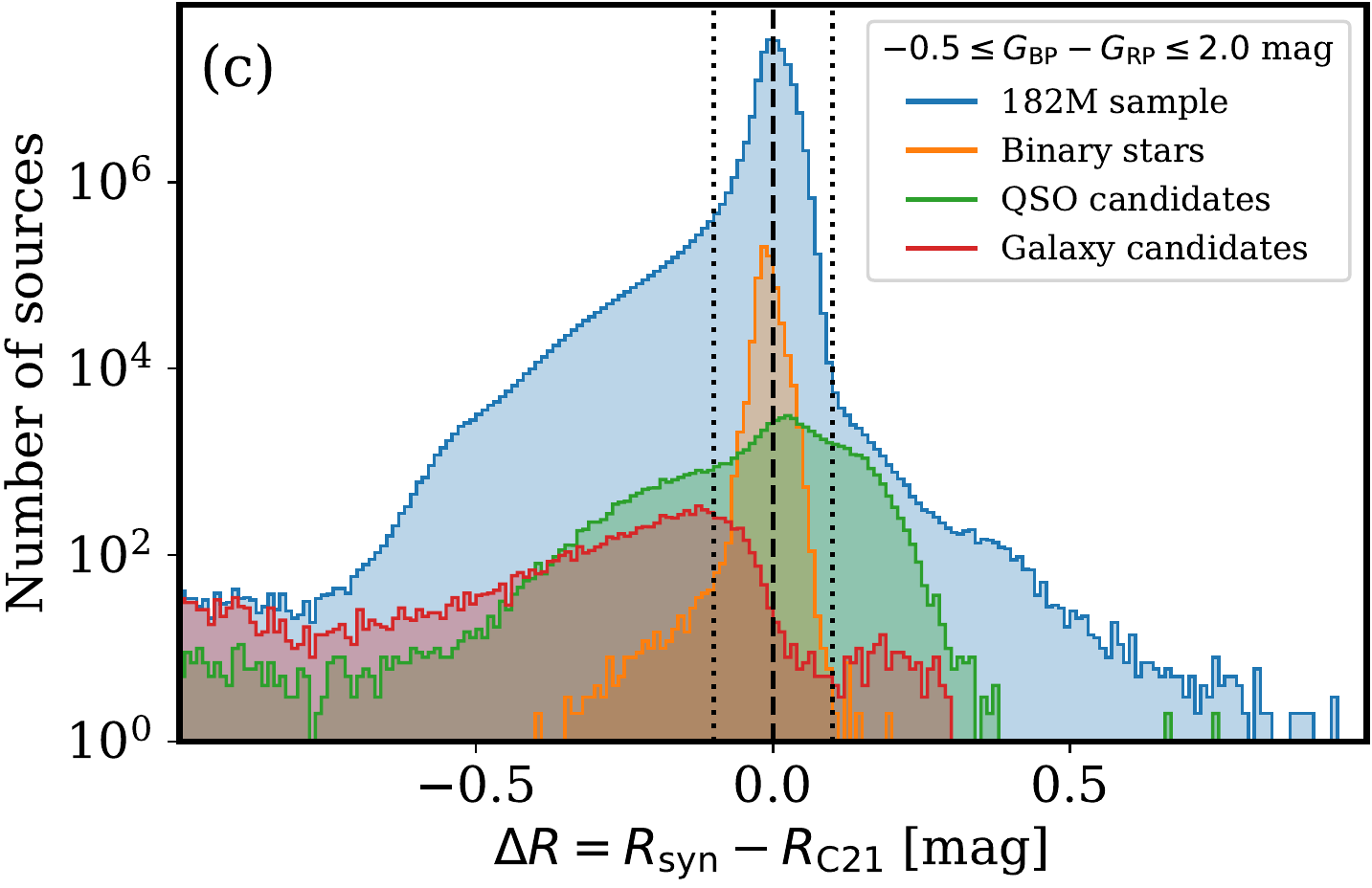}
    
    \caption{Histograms displaying the differences in the prediction of RGB magnitudes between the values derived in this work from~the {\gaia} XP low-resolution spectra and~those estimated using the polynomial functions published by C21. Since the latter are valid only for \mbox{$-0.5 \leq G_{\rm BP} - G_{\rm RP}\leq 2.0$ mag}, we applied here the same constraint to the 200 M sample (as indicated in the legend title), which reduces, in this analysis, the 200 M sample to \mbox{$\sim\!182$ million} objects (182~M sample). Panels~(\textbf{a}), (\textbf{b}) and~(\textbf{c}) show the results for the $B_{\rm RGB}$, $G_{\rm RGB}$ and R$_{\rm RGB}$ bands, respectively. Within~each panel, we  segregate the resulting histograms as shown in Figure~\ref{fig:hist_bp_rp}. The~vertical dashed line corresponds to \mbox{$\Delta {\rm X}={\rm X}_{\rm syn} - {\rm X}_{\rm C21}=0$~mag} with~\mbox{X\,=\,$B_{\rm RGB}$, $G_{\rm RGB}$ and $R_{\rm RGB}$}, whereas the vertical dotted lines are used in each panel to highlight  the $\pm 0.1$~mag interval, which encompasses 93.6\%, 97.8\% and~98.3\% of the sources in panels~(\textbf{a}), (\textbf{b}) and~(\textbf{c}), respectively. 2D~histograms representing the same dataset as a function of relevant parameters are shown in Figure~\ref{fig:map_errors_when_using_C21}.}
    \label{fig:histogram_errors_when_using_C21}
\end{figure}

Considering that the 200 million of sources with XP spectra in {\gdr3} represents only 10\% of the total two billion sources with available {\gaia} photometry, we explore here the validity of the polynomial calibrations published by C21 to estimate RGB magnitudes for the missing sources. With~this aim, we applied those polynomial functions to the 200 M subsample verifying \mbox{$-0.5 \leq G_{\rm BP}-G_{\rm RP}\leq 2.0$ mag} (the same constraint employed by C21) and~derived the differences with the RGB estimates derived from the synthetic photometry computed in this work. The~histograms of those differences are represented in Figure~\ref{fig:histogram_errors_when_using_C21} for the $B_{\rm RGB}$, $G_{\rm RGB}$ and~$R_{\rm RGB}$ bands (panels (a), (b) and~(c), respectively). 

Each panel represents the histogram corresponding to the full subsample after applying the colour constraint ($\sim 182$~million objects---hereafter, 182M samples; blue filled histogram) as~well as the histograms of those sources that are not classified as single stars (with the same criteria employed in Figure~\ref{fig:hist_bp_rp}). Interestingly, 93.6\%, 97.8\% and 98.3\% of the sources exhibit differences within the $\pm 0.1$~mag interval for $B_{\rm RGB}$, $G_{\rm RGB}$ and $R_{\rm RGB}$, respectively, (note that the vertical scale in the histograms is logarithmic). It is also evident that RGB predictions for objects flagged as {\texttt{in\_qso\_candidates} and \texttt{in\_galaxy\_candidates}} exhibit much larger deviations, although~they have a small contribution in the total number of~sources.

\begin{figure}[H]
    \includegraphics[width=0.32\textwidth]{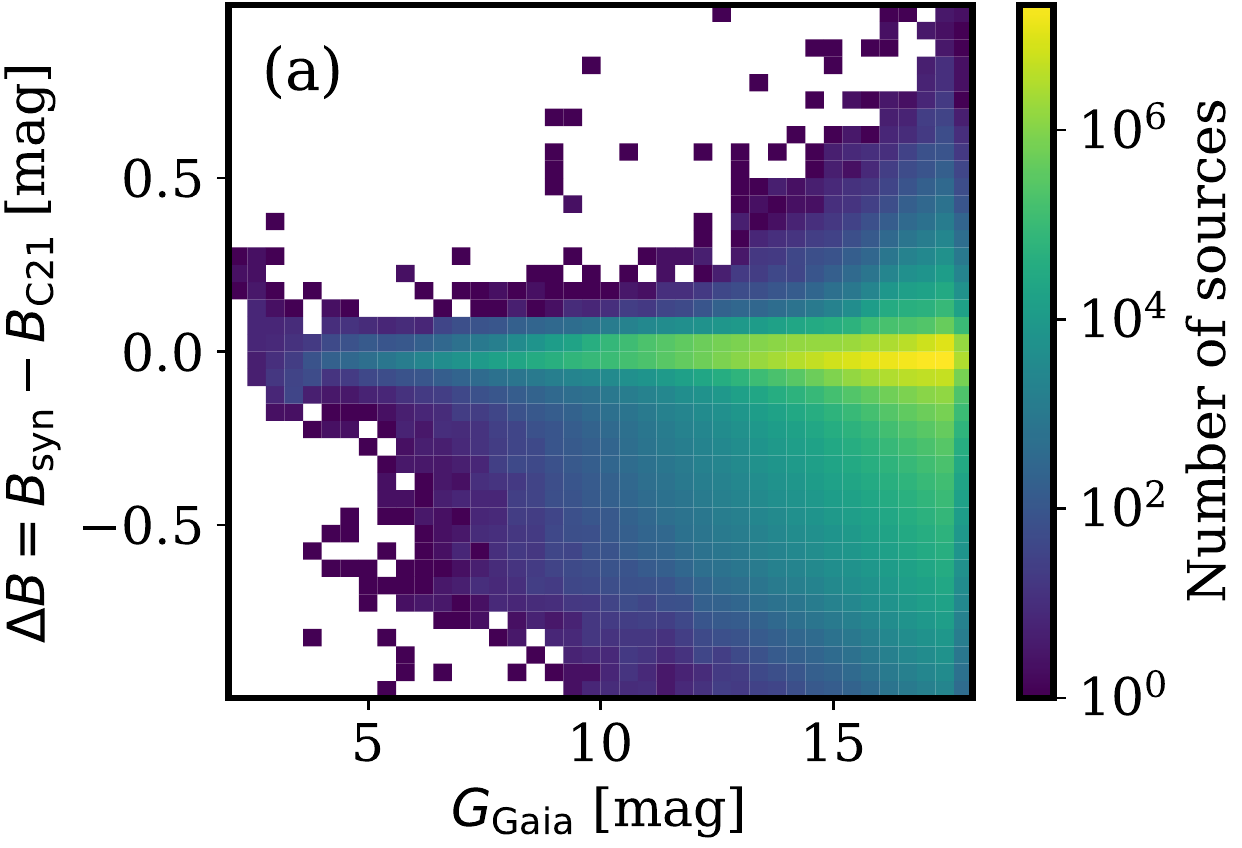}
    \hfill
    \includegraphics[width=0.32\textwidth]{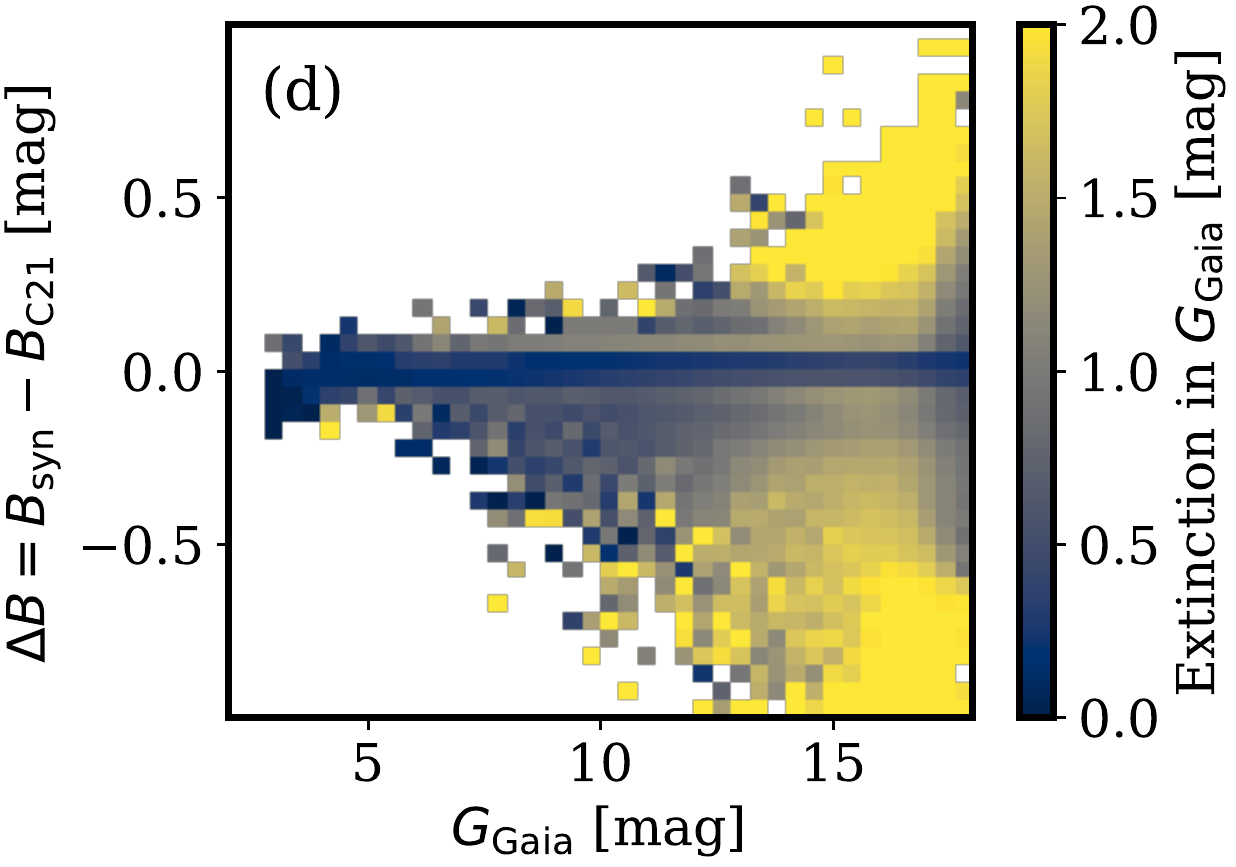}
    \hfill
    \includegraphics[width=0.32\textwidth]{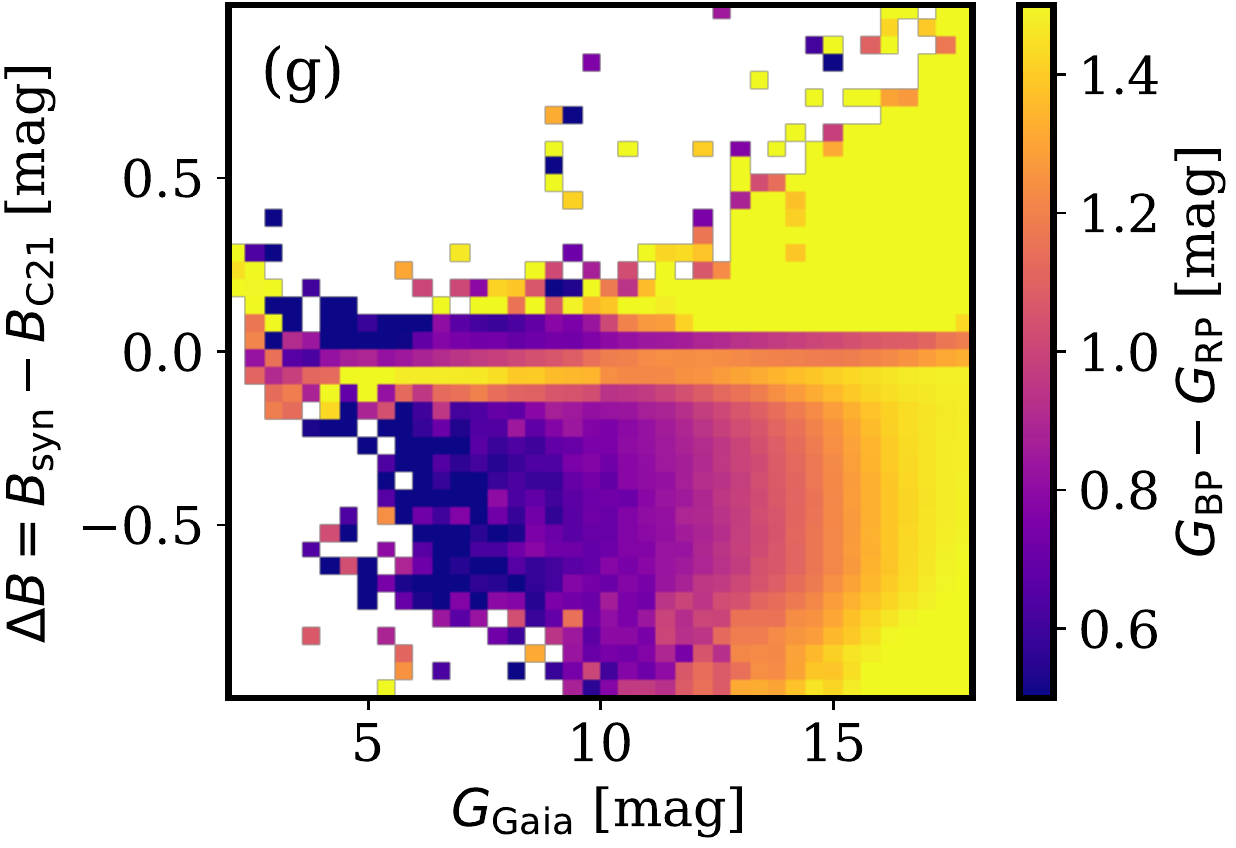}

    \bigskip
    
    \includegraphics[width=0.32\textwidth]{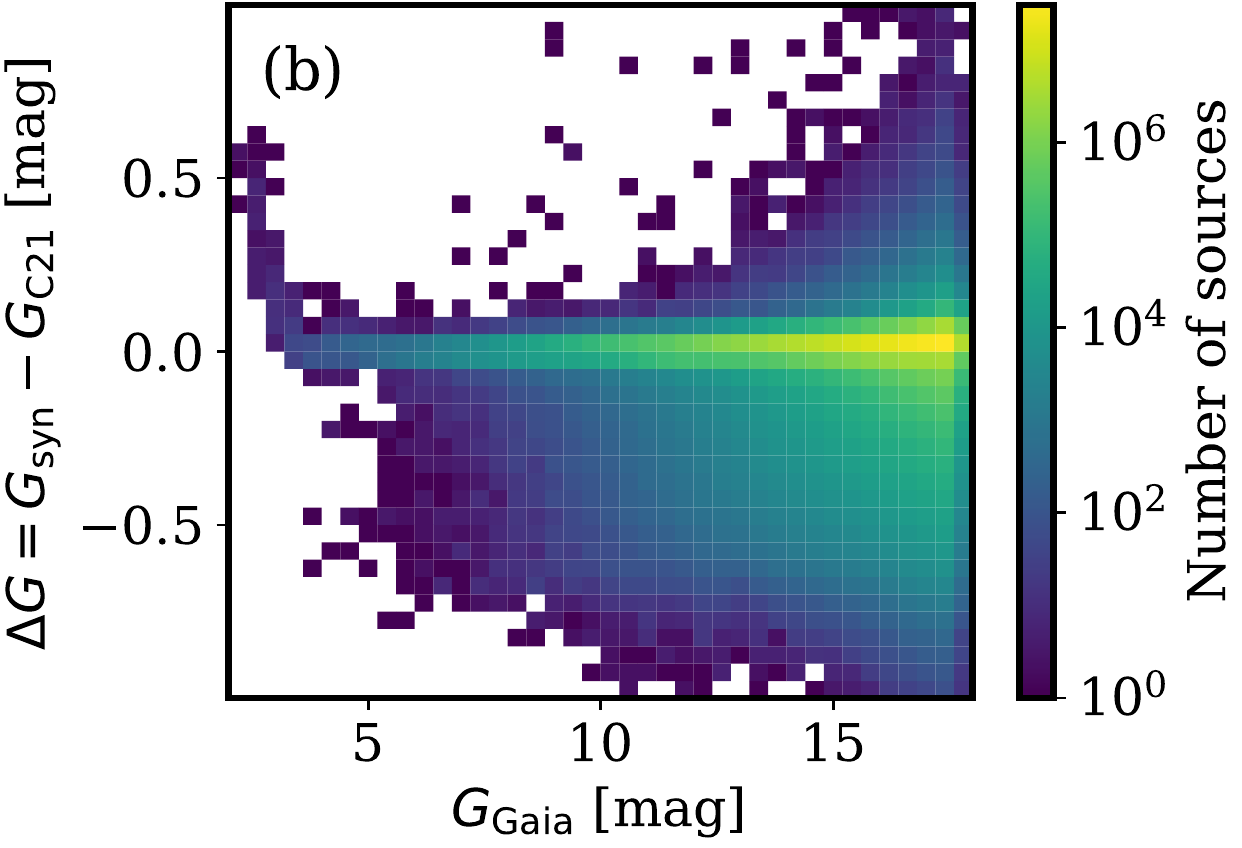}
    \hfill
    \includegraphics[width=0.32\textwidth]{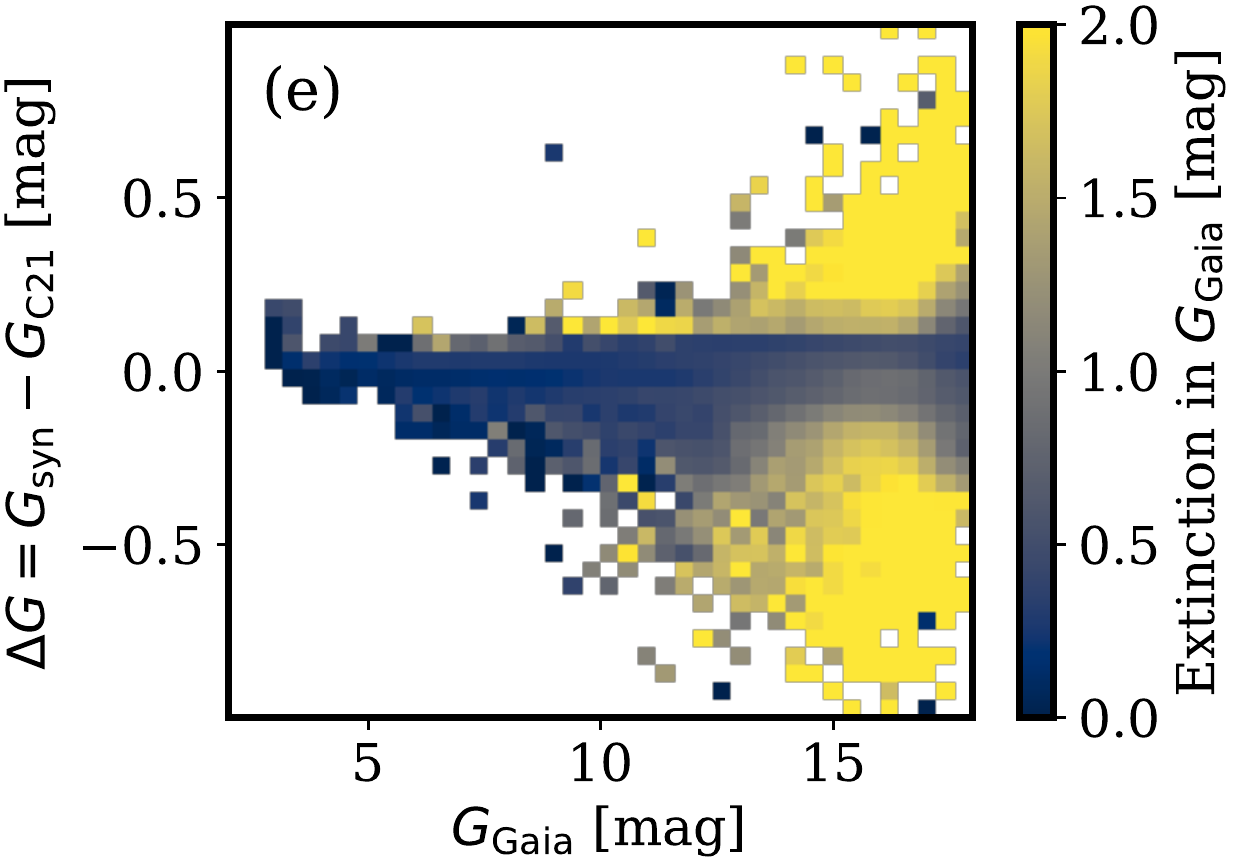}
    \hfill
    \includegraphics[width=0.32\textwidth]{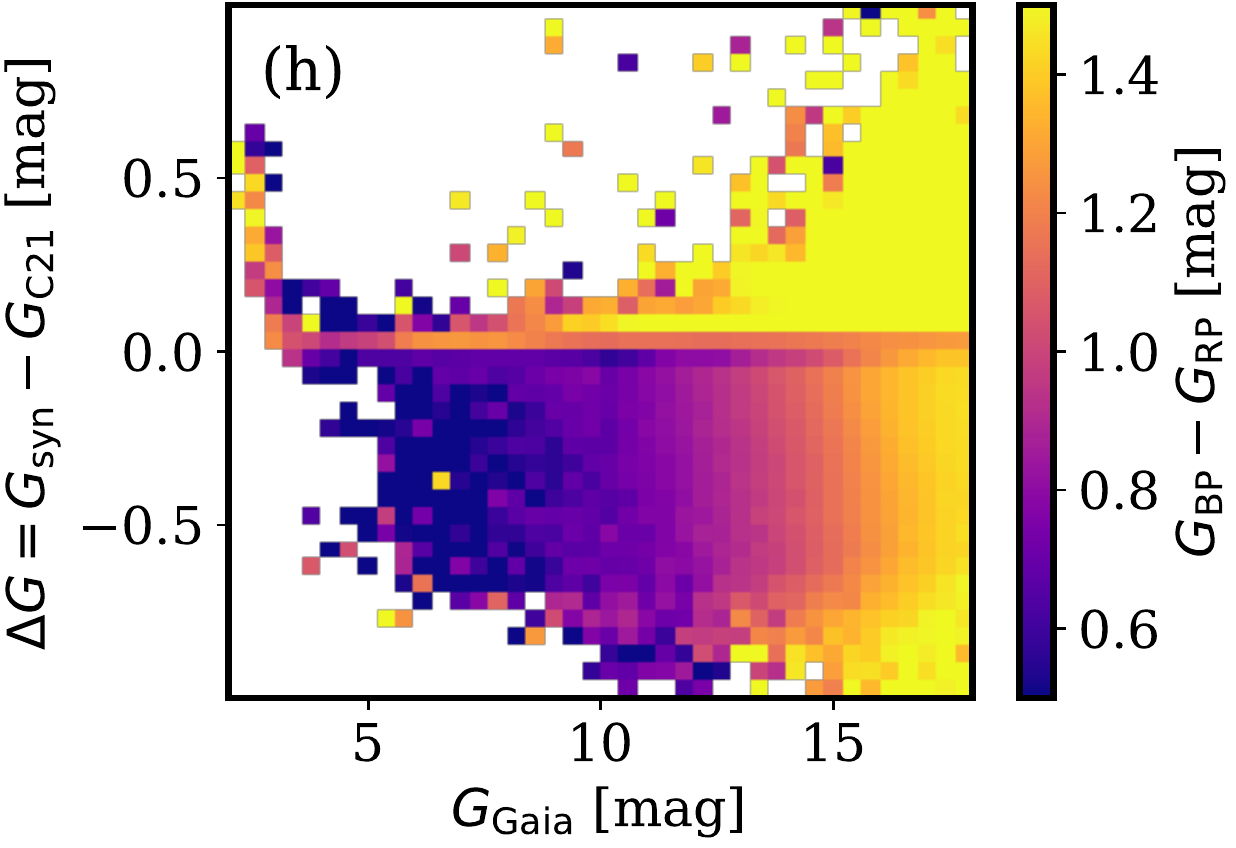}

    \bigskip
    
    \includegraphics[width=0.32\textwidth]{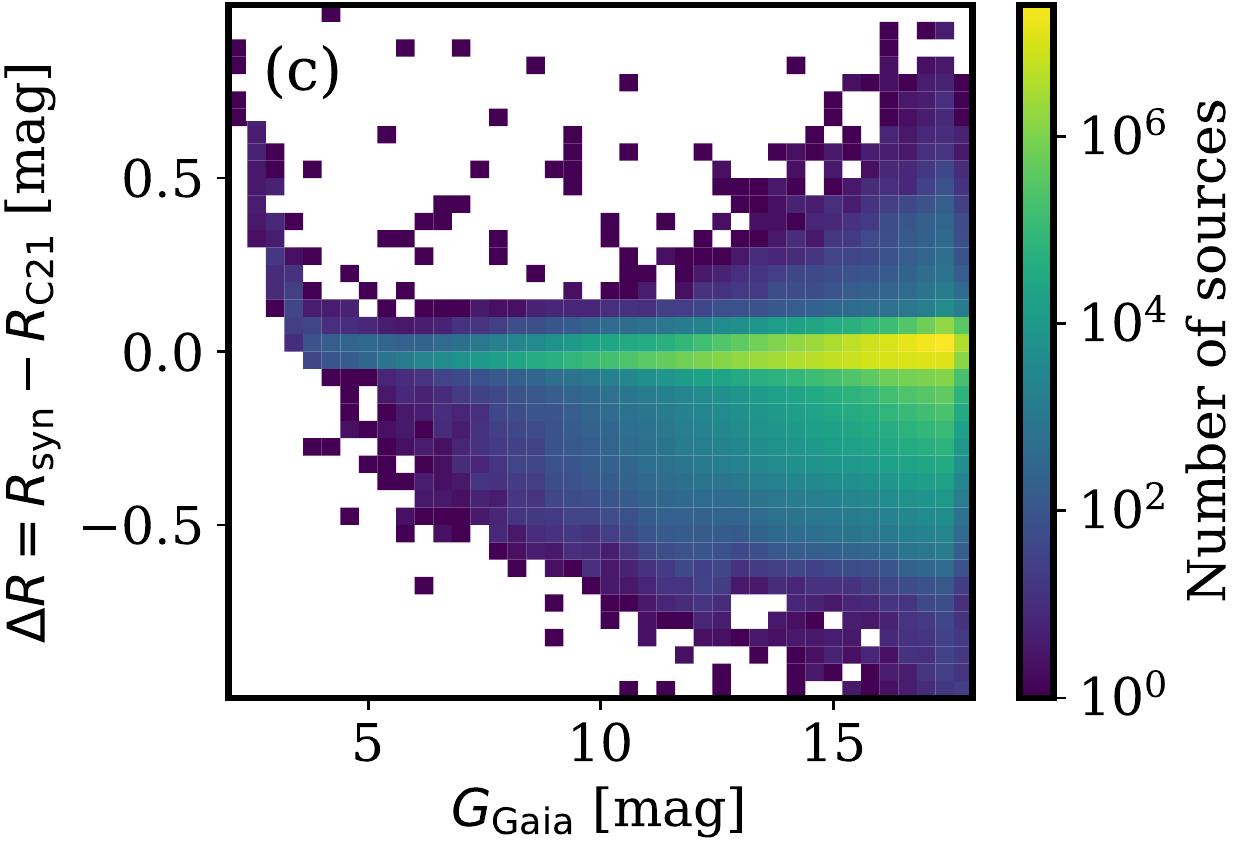}
    \hfill
    \includegraphics[width=0.32\textwidth]{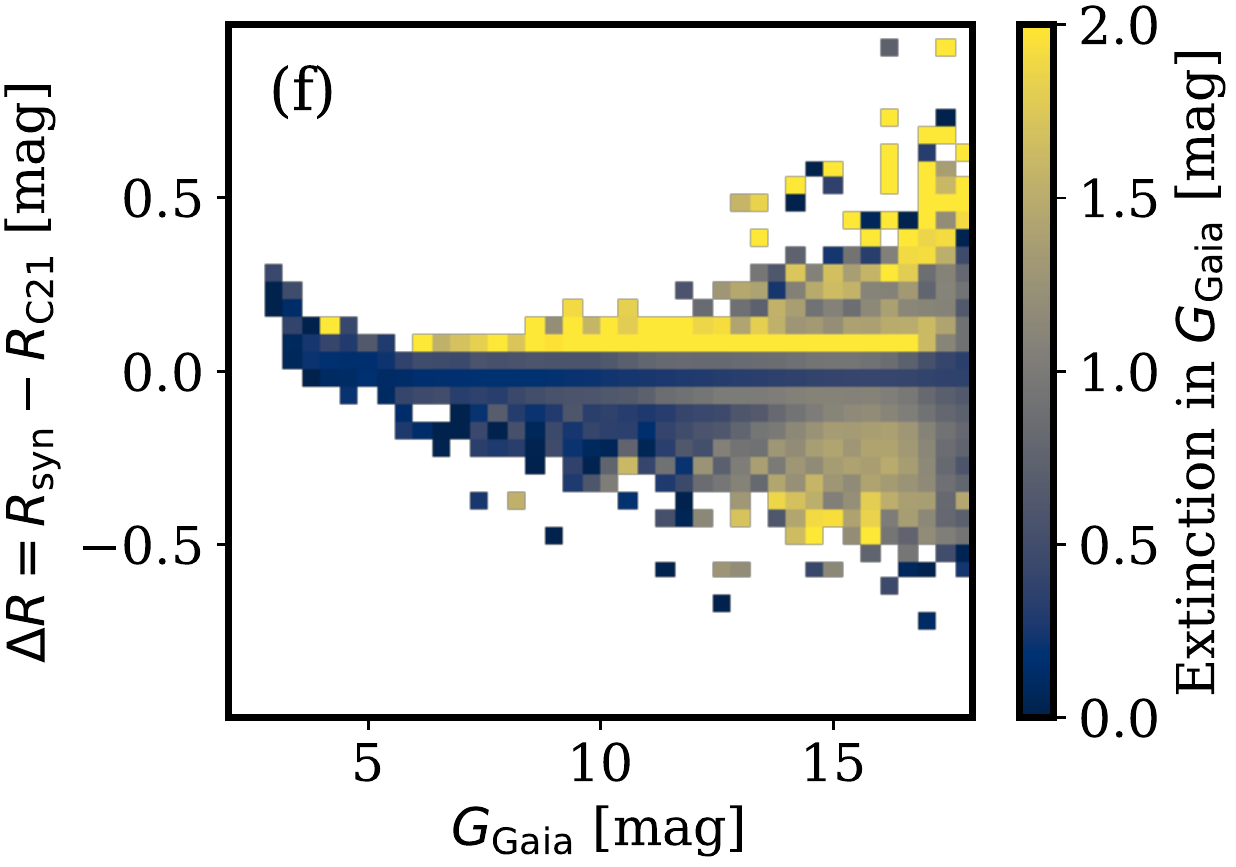}
    \hfill
    \includegraphics[width=0.31\textwidth]{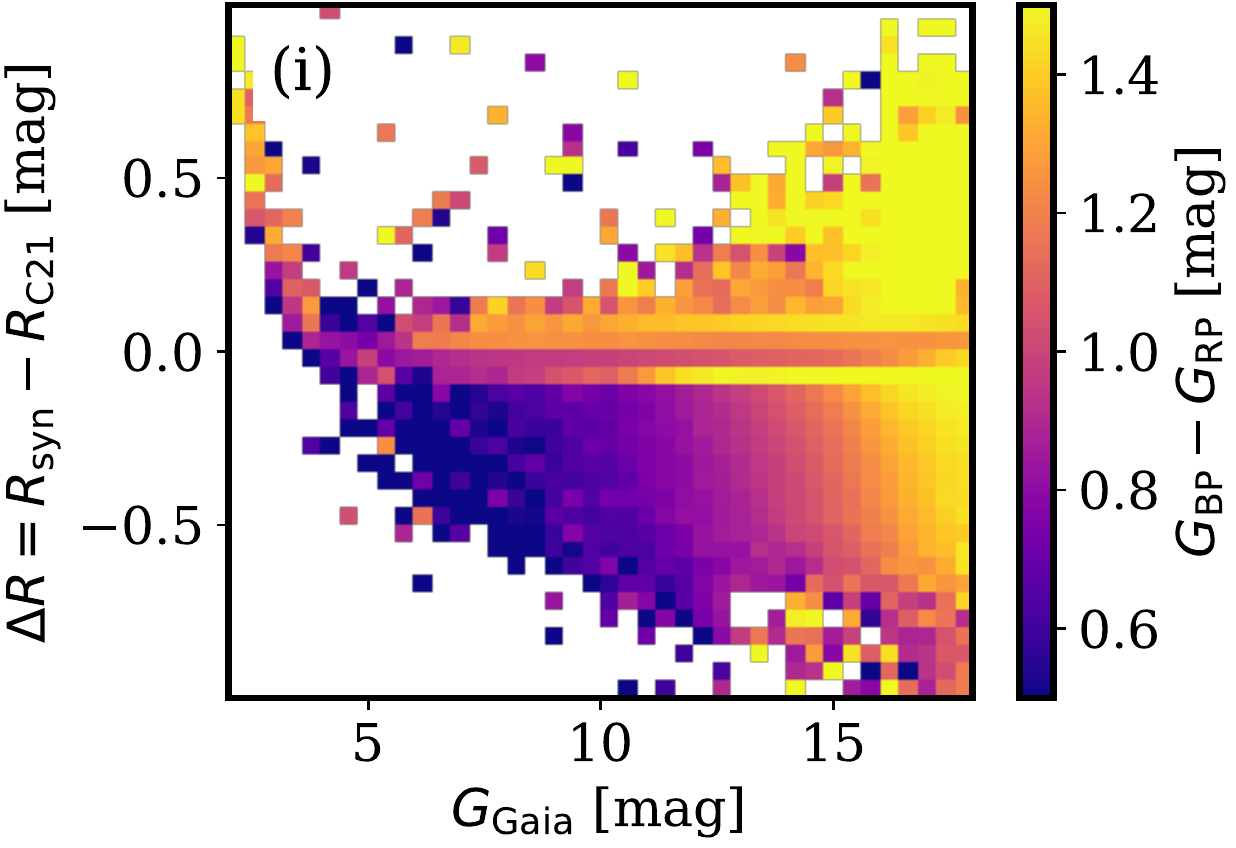}
    \caption{2D histograms representing the number of sources (left column, panels~(\textbf{a}--\textbf{c})), the~extinction in ${\rm G}_{\rm Gaia}$ (middle column, panels~(\textbf{d}--\textbf{f}))  and~the colour \mbox{${\rm G}_{\rm BP} - {\rm G}_{\rm RP}$} (right column, panels~(\textbf{g}--\textbf{i}))  as~a function of ${\rm G}_{\rm Gaia}$ (the horizontal axis; using 0.4~mag bins) and \mbox{$\Delta {\rm X}={\rm X}_{\rm syn} - {\rm X}_{\rm C21}=0$~mag}  with~\mbox{X\,=\,$B_{\rm RGB}$, $G_{\rm RGB}$ and $R_{\rm RGB}$} (the vertical axis; employing 0.05~mag bins). Each row displays the results for the $B_{\rm RGB}$, G$_{\rm RGB}$ and $R_{\rm RGB}$ bands (from top to bottom, respectively). These plots are an expanded version of the blue histograms shown in Figure~\ref{fig:histogram_errors_when_using_C21}, i.e.,~we are using the subset of the 200 M sample verifying \mbox{$-0.5 \leq {\rm G}_{\rm BP}-{\rm G}_{\rm RP}\leq 2.0$ mag}.}
    \label{fig:map_errors_when_using_C21}
\end{figure}

An expanded representation of the above results is shown in the 2D histograms displayed in Figure~\ref{fig:map_errors_when_using_C21}, where the magnitude differences are represented as a function of ${\rm G}_{\rm Gaia}$ with~colour coding for the histograms according to the source density (first column, panels (a), (b) and~(c)), extinction in $G_{\rm Gaia}$ (second column, panels (d), (e) and~(f)) and \mbox{$G_{\rm BP} - G_{\rm RP}$} (third column, panels (g), (h) and~(i)). These histograms reveal a systematic offset for very bright sources (saturated in {\gaia} XP spectra; see Riello et al.~\cite{Riello2021} and De Angeli et al.~\cite{DeAngeli2022}). 

The~figure also shows an expected increase in the differences of the magnitude predictions when considering fainter objects. Interstellar extinction has an important impact on the derived residuals, specifically in panels (d) and~(e) for sources with \mbox{$G_{\rm Gaia} > 13\;~{\rm mag}$}. There are also some clear systematic magnitude differences depending on the source \mbox{$G_{\rm BP} - G_{\rm RP}$} colour. Notwithstanding these relevant differences in the prediction of RGB magnitudes when using the C21 polynomial functions, it is important to highlight that, as~previously mentioned, the~predictions still fall within the $\pm 0.1\;{\rm mag}$ interval for a large fraction of the considered sources. 

Thus, although~the 200 M sample  provides  more reliable RGB predictions, the~C21 calibrations may still be useful when used with the corresponding caution (i.e., avoiding high extinction regions, restricting the \mbox{$G_{\rm BP} - G_{\rm RP}$} source colour and~using a large number of calibrating sources in order to derive statistical averages).

\unskip

\section{Accessing the 200 M~Sample}
\label{sec:rgbloom}

The 200 M sample produced in this work is accessible through an online table (Table~\ref{tab:RGBsyn} shows the first rows of the catalogue), which is available as a large, (5.0~GB) single, compressed CSV file (available at \url{https://nartex.fis.ucm.es/~ncl/rgbphot/gaiaDR3/RGBsynthetic_NOVARIABLES/sortida_XpContinuousMeanSpectrum_RGB_NOVARIABLES_final.csv.gz},
 accessed on 24 March 2023) (15~GB after decompressing). The~flag $-99.0$ is employed in this table to indicate missing values in any of the RGB magnitudes or in their associated~uncertainties.

As mentioned in Section~\ref{sec:methodology}, we have not removed the small fraction of objects in the 200 M sample classified in {\gdr3} as belonging to one of the following categories: \texttt{non\_single\_star} (0.31\% of the sample), \texttt{in\_qso\_candidates} (0.03\%) and~\texttt{in\_galaxy\_candidates} (0.01\%). Although~none of them have been thus far identified as a variable by {\gaia}, their use as reliable RGB calibrators should be properly checked through the statistical comparison with reliable sources located in their neighbourhood. All these sources are identified with the flag objtype in Table~\ref{tab:RGBsyn}.

\begin{table}[H]
\setlength{\tabcolsep}{2.7mm}
\caption{First rows of the online table with the synthetic RGB magnitudes ($B_{\rm RGB}$, $G_{\rm RGB}$ and $R_{\rm RGB}$) and their associated uncertainties ($\sigma_B$, $\sigma_G$ and $\sigma_R$)  derived for the 200 M sample from the {\gdr3} XP spectra for non-variable sources in the {\gaia} catalogue with id number equal to source\_id. 
{{The column objtype} is a flag that indicates the type of object in {\gdr3}: 1~for sources flagged as \texttt{non\_single\_star}, 2~for objects in \texttt{in\_qso\_candidates}, 3~for sources in \texttt{in\_galaxy\_candidates} and~0~for the rest of the sample. In~addition, the~last column, qlflag, provides a global quality flag, which is~0 for reliable sources and~~1 for objects with any indication of potential problem (blending, contamination or non-stellar identification).}
The full table can be downloaded from the link provided in the text,
and will also be available at the CDS.
}
\label{tab:RGBsyn}
\begin{tabularx}{\textwidth}{rcccccccc}
\toprule
\multicolumn{1}{c}{\textbf{source\_id}} 
&  \boldmath{$B_{\rm RGB}$} & \boldmath{$G_{\rm RGB}$} & \boldmath{$R_{\rm RGB}$} & \boldmath{$\sigma_B$} & \boldmath{$\sigma_G$} & \boldmath{$\sigma_R$} & {\bf objtype} & {\bf qlflag} \\
\midrule
4295806720 & 18.245 & 17.935 & 17.678 & 0.011 & 0.009 & 0.011 & 0 & 0 \\
38655544960 & 15.003 & 14.570 & 14.171 & 0.003 & 0.002 & 0.002 & 0 & 0 \\
1275606125952 & 16.849 & 16.519 & 16.266 & 0.005 & 0.004 & 0.006 & 0 & 0 \\
1653563247744 & 16.544 & 16.336 & 16.196 & 0.005 & 0.004 & 0.005 & 0 & 0 \\
2851858288640 & 12.779 & 12.550 & 12.387 & 0.001 & 0.001 & 0.002 & 0 & 0 \\
3332894779520 & 13.335 & 12.894 & 12.539 & 0.002 & 0.001 & 0.002 & 0 & 0 \\
3371550165888 & 15.314 & 14.938 & 14.615 & 0.003 & 0.002 & 0.003 & 0 & 1 \\
3508989119232 & 15.736 & 15.431 & 15.199 & 0.003 & 0.003 & 0.003 & 0 & 0 \\
4711579935744 & 14.736 & 14.481 & 14.300 & 0.002 & 0.002 & 0.003 & 0 & 0 \\
4814659150336 & 18.490 & 17.854 & 17.297 & 0.013 & 0.009 & 0.009 & 0 & 1 \\
5192616270720 & 18.370 & 17.551 & 16.964 & 0.013 & 0.007 & 0.007 & 0 & 0 \\
5291399870976 & 17.301 & 17.033 & 16.842 & 0.006 & 0.005 & 0.007 & 0 & 0 \\
5291399871488 & 18.935 & 18.296 & 17.686 & 0.019 & 0.012 & 0.011 & 0 & 0 \\
\multicolumn{1}{c}{$\cdot$} & $\cdot$  & $\cdot$ & $\cdot$ & $\cdot$ & $\cdot$ & $\cdot$ & $\cdot$ & $\cdot$ \\
\multicolumn{1}{c}{$\cdot$} & $\cdot$  & $\cdot$ & $\cdot$ & $\cdot$ & $\cdot$ & $\cdot$ & $\cdot$ & $\cdot$ \\
\multicolumn{1}{c}{$\cdot$} & $\cdot$  & $\cdot$ & $\cdot$ & $\cdot$ & $\cdot$ & $\cdot$ & $\cdot$ & $\cdot$ \\
\bottomrule
\end{tabularx}
\end{table}

\begin{figure}[H]
    \includegraphics[width=0.55\textwidth]{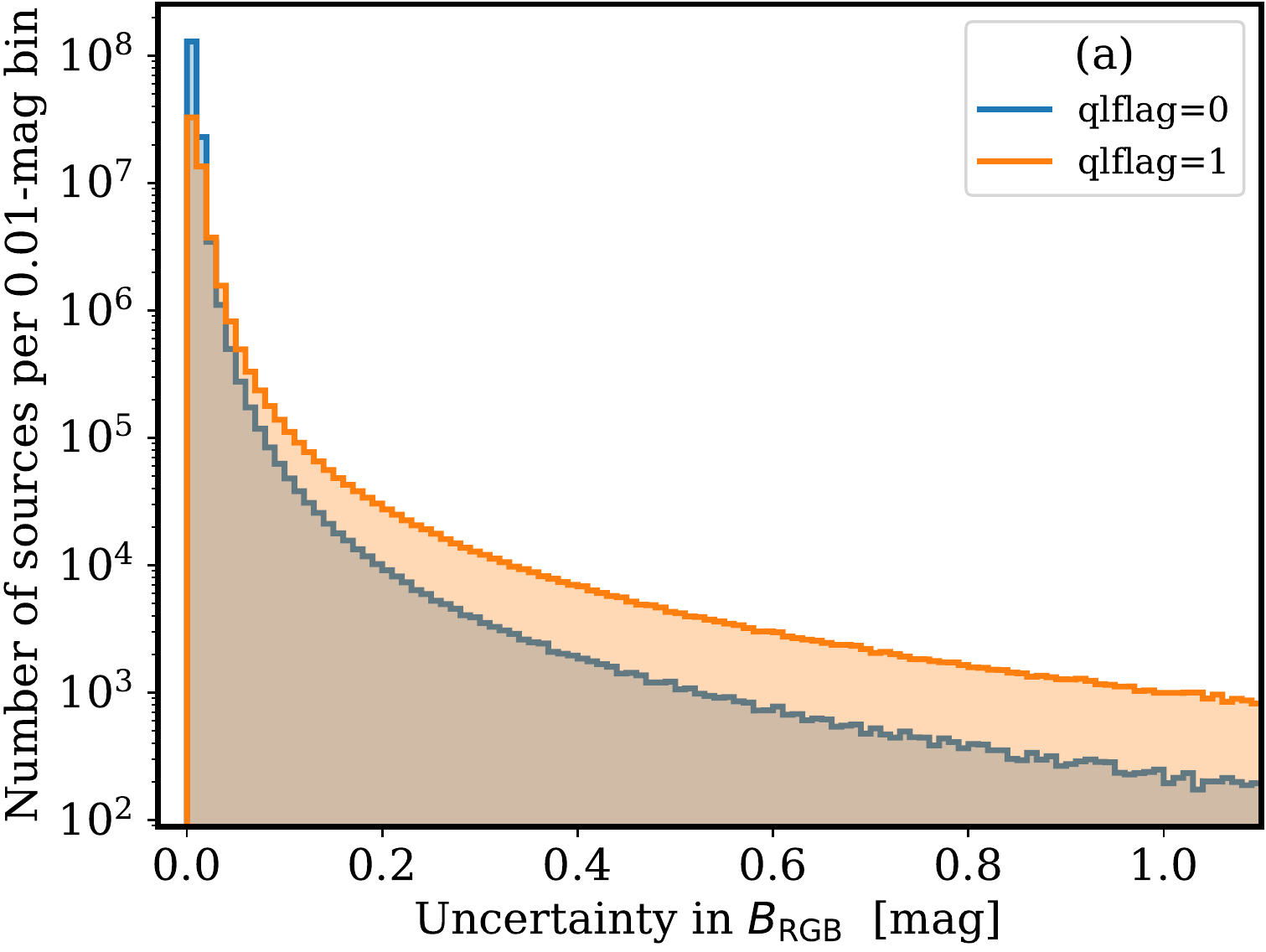}
    
    \bigskip
    
    \includegraphics[width=0.55\textwidth]{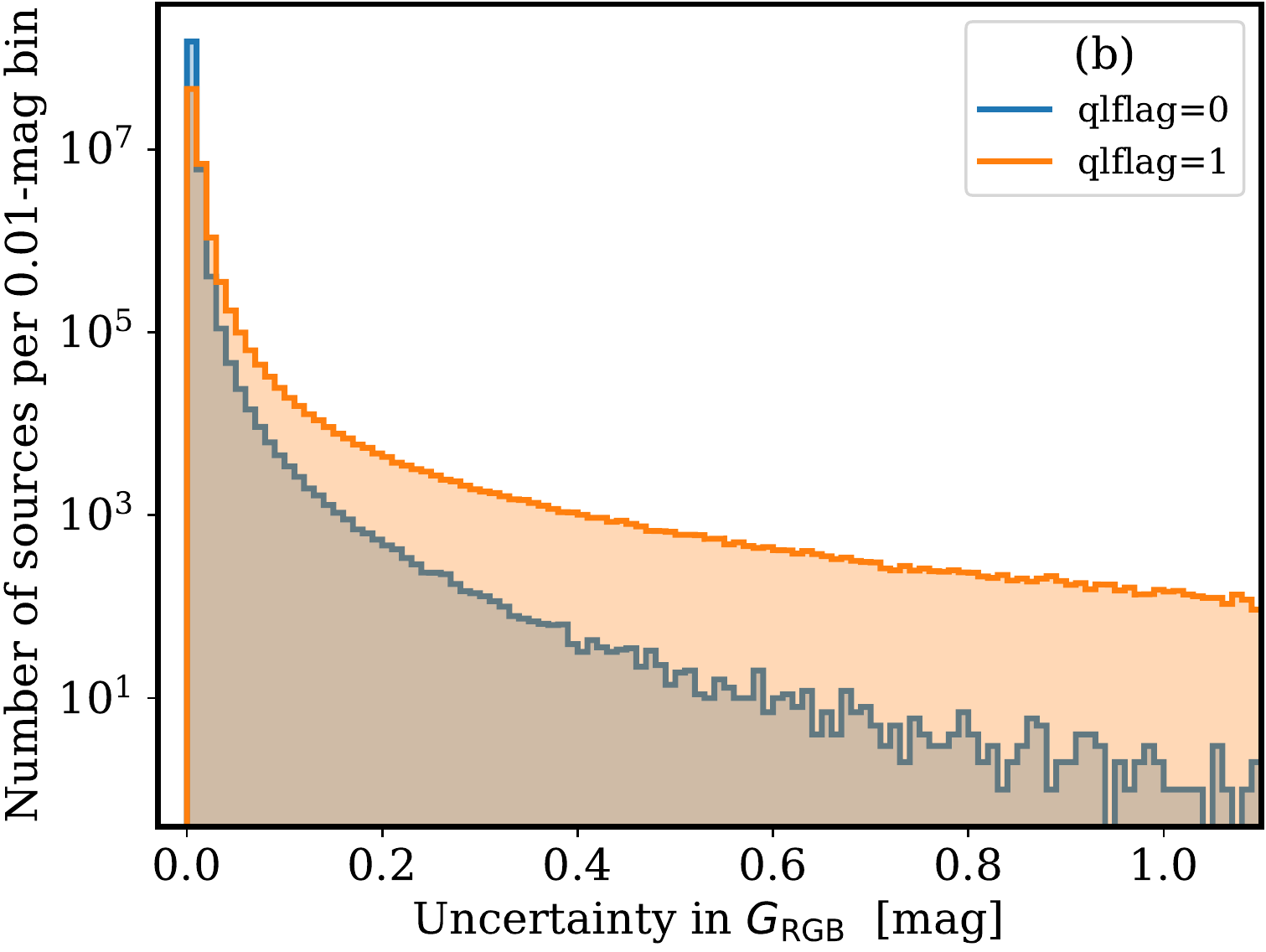}
    
    \bigskip
    
    \includegraphics[width=0.55\textwidth]{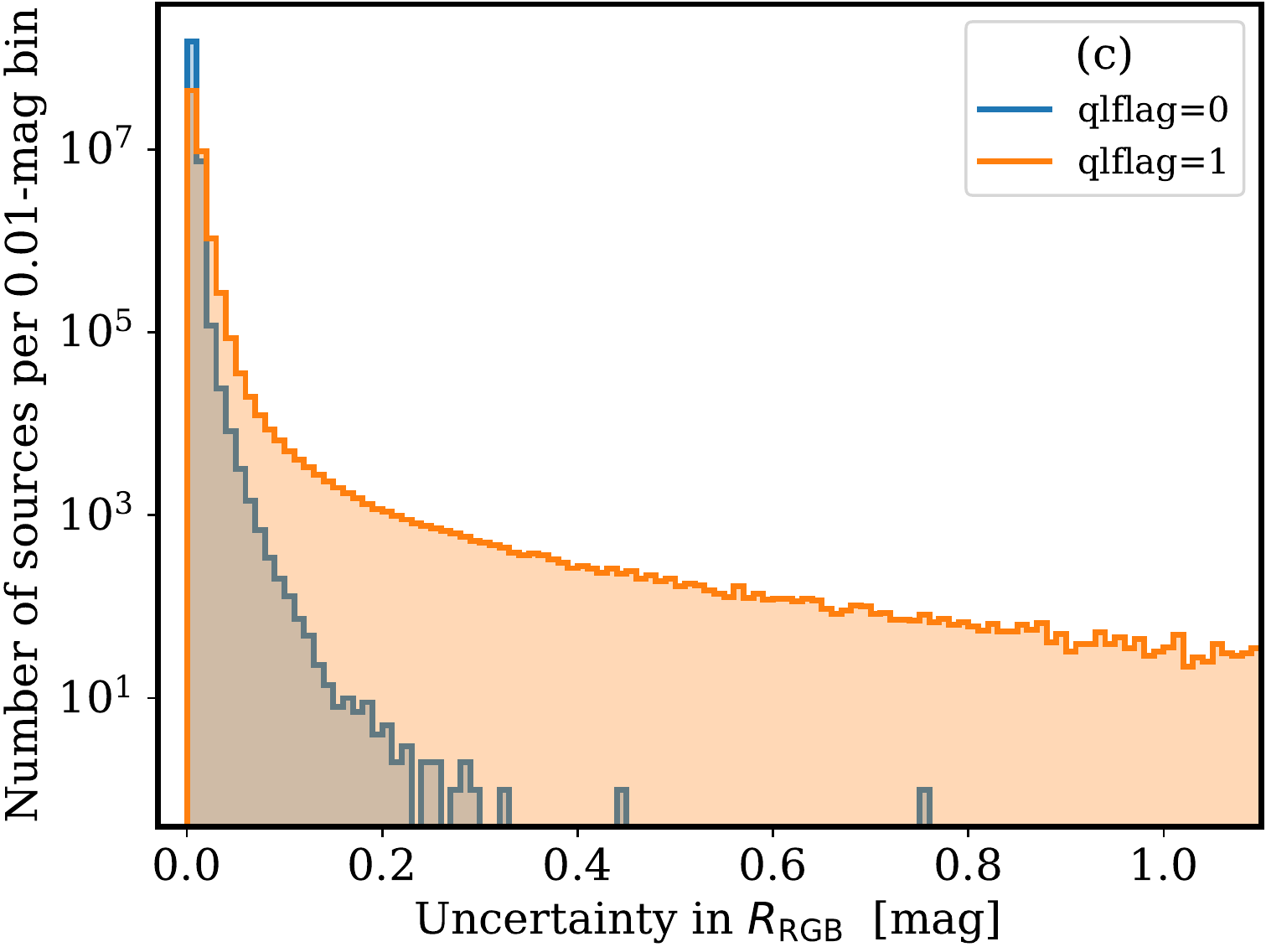}
    
    \caption{Histograms displaying the differences in the predicted uncertainties in the RGB synthetic magnitudes (panels~(a), (b) and (c) for $B_{\rm RGB}$, $G_{\rm RGB}$ and $R_{\rm RGB}$, respectively) as a function of the global quality parameter qlflag (last column in Table~\ref{tab:RGBsyn}). As~expected, the~objects with poorer quality (\mbox{qlflag = 1}) exhibit larger~uncertainties.}
    \label{fig:histogram_quality_errors}
\end{figure}

In addition, we also assigned a global quality flag to each source, qlflag, provided in the last column of Table~\ref{tab:RGBsyn}). We define \mbox{qlflag = 0} for reliable sources (74.2\% of the 200 M sample) and~assign \mbox{qlflag = 1} for those objects suspicious of suffering any potential problem (blending, contamination or non-stellar identification; 25.8\% of the sample). As~expected, the~uncertainties in the RGB synthetic magnitudes are larger for the \mbox{qlflag = 1} sources as~displayed in Fig.~\ref{fig:histogram_quality_errors}.

Complementary to the online table, in~order to ease  access to  synthetic RGB photometry, we created a Python package called \mbox{\sc rgbloom} (available at \url{https://github.com/guaix-ucm/rgbloom}, accessed on 24 March 2023), which performs cone search queries around any position on the celestial sphere. This code is an updated version of \mbox{\sc rgblues} (available at \url{https://github.com/guaix-ucm/rgblues}, accessed on 24 March 2023), the~tool created by C21 to retrieve RGB magnitudes from the 15~M~sample.

Once installed, the~code can be easily executed from the command line---for~example, using a command like this one:

\texttt{\$ rgbloom 56.66 24.10 1.0 12}

The four numerical arguments correspond to the position search values (RA, DEC and search radius) in decimal degrees and the~limiting $G_{\rm Gaia}$~magnitude. In~the example command shown here, we are searching for sources brighter that $G_{\rm Gaia}=12$~mag within~a circular aperture of radius 1~degree with~centre at right ascension \mbox{RA = 56.66~deg} and declination \mbox{DEC=24.10~deg} (corresponding to the Pleiades star cluster).

\begin{figure}[H]
    \includegraphics[width=\textwidth]{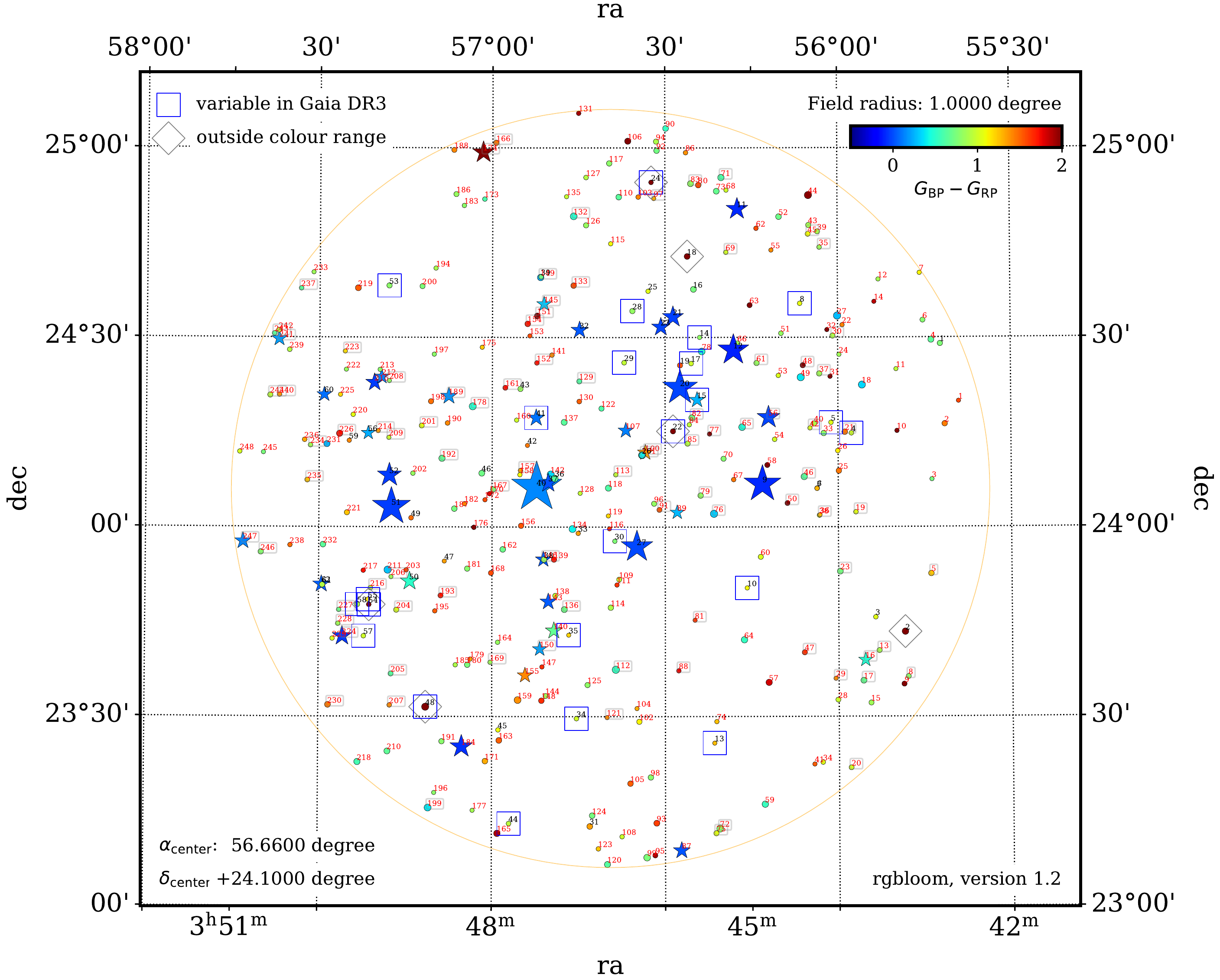}
    \caption{Example of a finding chart generated by the Python package {\sc rgbloom} after performing a cone search centred in the Pleiades star cluster with~a search radius of \mbox{1~degree}. The~objects in this plot are colour coded based on the {\gaia} ${\rm G}_{\rm BP}-{\rm G}_{\rm RP}$ colour and~are numbered with labels of different colours (red or black for objects belonging or not to the 200 M sample, respectively) with~numbers matching the first column of the output files \texttt{rgbloom\_200m.csv} and \texttt{rgbloom\_no200m.csv} generated during the execution of the code. The identification number
 of the less reliable sources in {\texttt{rgbloom\_200m.csv}} (those with \mbox{qlflag=1}) appear within a rectangle with a light-gray colour. For sources that do not belong to the 200 M sample, and where RGB estimates correspond to the polynomial transformations from C21, which can be affected by systematic biases, the~program overplots a blue square when the {\gaia} DR3 {\texttt{phot\_variable\_flag}} is set to {\texttt{VARIABLE}} and~a grey diamond when the source colour is outside the \mbox{$-0.5 \leq {\rm G}_{\rm BP}-{\rm G}_{\rm RP}\leq 2.0$ mag} interval.}
    \label{fig:rgbloom_pleiades}
\end{figure}

The steps followed by {\sc rgbloom} to provide its output are the following:

\begin{enumerate}

\item Cone search in {\gdr3} down to the pre-defined limiting  $G_{\rm Gaia}$~magnitude, retrieving the following parameters: \texttt{source\_id}, \texttt{ra}, \texttt{dec},
\texttt{phot\_g\_mean\_mag}, \texttt{phot\_bp\_mean\_mag},
\texttt{phot\_rp\_mean\_mag} and \texttt{phot\_variable\_flag}.

\item Initial RGB magnitude estimation using the polynomial transformations given in \mbox{Equations~(2)--(4)} by C21. 

\item Retrieval of the RGB synthetic magnitudes for sources in the 200 M sample within the {\sc healpix} level-8 tables enclosing the region of the sky defined in the initial cone~search.

\item Cross-matching between the {\gdr3} and the 200 M subsamples to identify sources with RGB synthetic magnitudes estimated from the XP low-resolution~spectra.

\item Generation of the output files. In~particular, two files (in CSV format) are generated: \texttt{rgbloom\_200m.csv}, which contains the sources belonging to the 200 M sample with RGB synthetic magnitudes; and \text{rgbloom\_no200m.csv}, which includes sources that do not belong to the 200 M sample. In~this latter case, the~RGB magnitudes provided in the file correspond to the estimates derived  using the polynomial relationships in C21. It is important to remember (see Section~\ref{sec:comparison_with_C21}) that these estimates are more uncertain than the new RGB values computed in this work and that they can be biased due to systematic effects introduced by interstellar extinction by~exhibiting a colour outside the \mbox{$-0.5 \leq G_{\rm BP}-G_{\rm RP} \leq 2.0$} interval (where the C21 calibrations were computed) or~by variability of the~source.

\item Creation of a finding chart of the results (see the example in Figure~\ref{fig:rgbloom_pleiades} for the region around the Pleiades star cluster). The~users of {\sc rgbloom} should rely more strongly on the RGB estimates corresponding to sources belonging to the 200 M sample (labelled with red numbers in Figure~\ref{fig:rgbloom_pleiades}) and~make judicious use of the predictions that rely on the C21 polynomial calibration (labelled with black numbers in the same figure) as~discussed in Section~\ref{sec:validity_C21_polynomial_calibration}. Nevertheless, since the sky coverage of the 200 M sample is still not very good at certain high Galactic latitudes (see Figure~\ref{fig:mollweide_maps}), the~RGB estimates from the C21 polynomial calibration may still be useful after discarding the sources with large interstellar~extinction.

\end{enumerate}

\section{{Conclusions}}
\label{sec:conclusions}

We provided, in this work, synthetic photometry in RGB passbands derived from {\gdr3} XP spectrophotometry. The~RGB magnitudes derived in this work represent a significant improvement over those published by C21 as they were directly derived from space-based SED observations and~not predicted through photometric relationships as~in C21. Although~it is true that the {\gaia} XP spectra are of low spectral resolution by astrophysical standards, they constitute a clear improvement over the predictions obtained by the {\gedr} $G_{\rm Gaia}$, $G_{\rm BP}$ and~$G_{\rm RP}$ integrated photometry employed by C21 to derive their polynomial calibration. 
 
 In~addition, the~new RGB estimates were derived employing the {\gdr3} XP spectra, directly using their XP coefficients and their associated basis functions, which allows for preservation of the maximum information contained in the {\gaia} spectra as~well as confident estimation of the propagation of uncertainties in the derived~results.
The number of calibrated sources increased significantly, passing from $\sim 15$~million sources in C21 to slightly more that 200~million in this work. The~new 200 M sample can now include sources without extinction restrictions and~does not rely on any approximate calibration only valid for isolated solar metallicity~stars. 

In addition, the~RGB magnitude estimates are provided with their associated uncertainties, which, in the C21 sample, were roughly estimated to be within a $\pm 0.1$~mag interval and~ are now more robustly determined. In addition, the~uncertainties for a considerable number of sources were within $\pm 0.01$~mag, which constitutes a significant improvement. This means that, contrary to what happened with the C21 sample, whose RGB estimates should not be considered to be extremely accurate on a star-by-star basis, the~availability of reliable uncertainties in the new 200 M catalogue allows us to infer quality photometric calibration even with a very small number of reference~sources.

This work demonstrates that RGB photometry can already be performed using a vast catalogue of reliable calibration sources, available in a wide range of magnitudes and for a significant fraction of the celestial sphere. The~astronomical magnitudes can easily be transformed into radiometric units by using the formulas published in~\cite{sanchez2017sky}.

\vspace{6pt} 



\authorcontributions{Conceptualization, J.M.C., A.S.d.M., J.Z. and N.C.; Computation J.M.C., N.C., S.P. and R.G.; Data analysis on the NightUp database J.M.C., N.C., S.P. and R.G.; Field test A.S.d.M., J.I., J.Z., J.M.C. and E.M. All authors have read and agreed to the published version of the~manuscript.}

\acknowledgments
{This work was (partially) funded by the Spanish MICIN/AEI/10.13039/501100011033 and by ``ERDF A way of making Europe'' by the European Union through grants RTI2018-095076-B-C21 and PID2021-122842OB-C21 and~the Institute of Cosmos Sciences University of Barcelona (ICCUB, Unidad de Excelencia ’Mar\'{\i}a de Maeztu’) through grant CEX2019-000918-M. Funding for the DPAC
has been provided by national institutions---in~particular, the institutions
participating in the {\it Gaia} Multilateral Agreement.
This project  received funding from the European Union’s Horizon 2020 research and innovation programme under the Marie Skłodowska-Curie grant agreement No 847635 (UNA4CAREER). RALAN map project. 
This work  made use of data from the European Space Agency (ESA) mission
{\it {Gaia}} (\url{https://www.cosmos.esa.int/gaia}, accessed on {12 July 2022})  processed by the {\it Gaia}
Data Processing and Analysis Consortium ({DPAC,} \url{https://www.cosmos.esa.int/web/gaia/dpac/consortium}, accessed on {12 July 2022}). 
This work was possible thanks to the extensive use of IPython and Jupyter notebooks \citep{PER-GRA:2007}. This research made use of {\sc {astropy}} (\url{http://www.astropy.org}, accessed on {24 March 2023}), a~community-developed core Python package for Astronomy \citep{astropy:2013, astropy:2018}, {\sc numpy} \citep{harris2020array}, {\sc scipy} \citep{2020SciPy-NMeth}, {\sc matplotlib} \citep{4160265}, {\sc pandas} \citep{mckinney-proc-scipy-2010} and~{\sc vaex}~\citep{2018A&A...618A..13B}. Some of the results in this paper were derived using the healpy and HEALPix packages \citep{2005ApJ...622..759G, Zonca2019}.}

\dataavailability{The BP/RP coefficients for all sources used here were obtained from the {\gaia} {catalogue} (\url{https://gea.esac.esa.int/archive/}, accessed on 12 July 2022). From~them, we derived the RGB synthetic photometry,  which can be accessed by following the detailed instructions in Section~\ref{sec:rgbloom}.} 


\conflictsofinterest{ A. S\'anchez de Miguel discloses that he does consulting work sporadically for Savestars Consulting S.L. and is a member of the board of Cel Fosc. The~authors are not aware of any affiliations, memberships, funding or~financial holdings that might be perceived as affecting the objectivity of this~article.}



\begin{adjustwidth}{-\extralength}{0cm}

\reftitle{References}

\end{adjustwidth}
\end{document}